\documentclass[12pt]{article} 
\usepackage{graphicx}
\makeindex

\newcommand{\half}{0.45\textwidth}
\begin{document}

\title{\mbox{ }\\[-2.5cm] Micro- and Macrosimulation of Freeway Traffic}
\author{\normalsize Dirk Helbing$^{1,2}$, Ansgar Hennecke$^{1}$,
Vladimir Shvetsov$^{1}$, and Martin Treiber$^{1}$\\[2mm]
\normalsize $^1$II. Institute of Theoretical
Physics,  University of Stuttgart,\\ 
\normalsize Pfaffenwaldring 57/III, D-70550 Stuttgart, Germany\\[2mm]
\normalsize $^2$ Collegium Budapest~-- Institute for Advanced Study,\\
\normalsize Szenth\'{a}roms\'{a}g utca 2, 
H-1014 Budapest, Hungary} 
\maketitle              
\begin{abstract}
We present simulations of congested traffic
in circular and open systems with a non-local, gas-kinetic-based
traffic model and a novel car-following model.
The model parameters are all intuitive and can be easily
calibrated. Micro- and macrosimulations with these models for
identical vehicles on a single lane produce the same traffic 
states, which also qualitatively agree with empirical traffic
observations. Moreover, the phase diagrams of traffic states
in the presence of bottlenecks for
the microscopic car-following model and the macroscopic gas-kinetic-based
model almost agree. In both cases, we found metastable regimes, spatially
coexistent states, and a small region of tristability. The distinction of
different types of vehicles (cars and long vehicles) yields additional
insight and allows to reproduce empirical data even
more realistically, including the observed fluctuation properties
of traffic flows like the wide scattering of congested traffic data. 
\par
Finally, as an alternative to the gas-kinetic approach,
we propose a new scheme for deriving non-local
macroscopic traffic models from given microscopic car-following models. 
Assuming identical (macroscopic) initial and boundary conditions,
we show that there are microscopic 
models for which the corresponding macroscopic 
version displays an almost identical dynamics. This enables
us to combine micro- and macrosimulations of road sections by simple 
algorithms, and even to simulate them simultaneously.\\[2mm]
\end{abstract}

\section{Introduction}

In recent years, the community of scientists engaged in traffic modelling 
is rapidly growing (for overviews see \cite{Leu88,book,traf}). 
This is not only due to practical implications
for optimizing freeway traffic \cite{Bovy},
but also because of the observed non-equilibrium phase transitions 
\cite{kerner-sync,helb-nat} 
and the various non-linear dynamical phenomena like the formation of
traffic jams \cite{kk-93}, 
stop-and-go traffic \cite{Kue87,Kue84,kk-94}, 
and ``synchronized'' congested traffic 
\cite{kerner-rehb96-2,kerner-sync,HT-sync}.
It seems that all forms of congested traffic have
almost universal properties which are largely
independent of the initial conditions and the spatially averaged density, like
the characteristic outflow $Q_{\rm out}$ from traffic jams of about
$1800\pm 300$ vehicles per hour and lane or their typical dissolution 
velocity $C$ of about $-15\pm 5$ kilometers per hour
\cite{kerner-rehb96}. This universality arises from the
highly correlated state of motion produced by traffic
congestions. In particular, the
outflow $Q_{\rm out}$ is related to the time gap between successive
departures from the traffic jam \cite{kerner-chania,Ref}. 
Therefore, the outflow is practically
independent of the initial conditions and the kind of congested traffic.
As a consequence of the constant outflow, the propagation velocity
$C$ of jam fronts, given by the dissolution speed of traffic jams,
is nearly constant as well. 
\par
From a  physics point of view, a model for a real system (here: for 
unidirectional freeway traffic) should be as simple as possible, but
not simpler (i.e., parameter dependencies that are known to exist,
should be reflected by the model). The parameters of the model
should be intuitive,
easy to calibrate, and the corresponding values should be realistic.
Moreover, the traffic model 
should reproduce all observed
localized and extended traffic states \cite{Lee-emp},
including synchronized traffic and the wide scattering
of congested traffic data \cite{kerner-rehb96-2}.
Furthermore, the observed
hysteresis effects \cite{Treiterer,OCT}, complex states
\cite{kerner-rehb96-2,Kerner-wide},
and the existence of self-organized 
quantities like the constant propagation velocity of stop-and-go waves
or the outflow from a traffic jam \cite{kk-94} 
should be reproduced.
Finally, the dynamics must not lead to vehicle collisions
or exceed the maximum vehicle density,
and the model should allow for a fast numerical simulation.

In this paper, we discuss some of these criteria for the recently
proposed non-local, gas-kinetic-based traffic model (GKT model) 
\cite{HT-sync,GKT}
and the microscopic intelligent-driver model (IDM) \cite{coexist}. 
In particular, we discuss several different kinds of bottlenecks
and present the phase diagram of congested traffic states for open,
inhomogeneous systems \cite{Phase}.
We show that, for identical vehicles,
the macroscopic dynamics is qualitatively the same for both models.
Finally, we investigate heterogeneous (mixed multi-class traffic 
composed of cars and lorries/long vehicles) 
and propose a natural explanation of 
the observed wide scattering of congested traffic data.

\section{Macroscopic Traffic Models} \label{S2}

In contrast to microscopic traffic models, which delineate 
the positions $x_\alpha(t)$ and velocities $v_\alpha(t)$
of all interacting vehicles $\alpha$, 
macroscopic traffic models restrict to the description of 
the collective vehicle dynamics in terms of the spatial vehicle
density $\rho(x,t)$ and the average velocity $V(x,t)$ as a function of
the freeway location $x$ and time $t$. One advantage of macroscopic
traffic models is that they allow to simulate the traffic dynamics in several
lanes by effective one-lane models considering a certain probability
of overtaking. In the following, we will use this approach, although
it is also possible to develop macroscopic traffic models for several
lanes with explicit treatment of lane-changing, as mentioned
later on.
\par
Being a consequence of the conservation of the number of vehicles,
virtually all macroscopic traffic models are based on the continuity equation 
\begin{equation}
 \frac{\partial \rho}{\partial t}
 + \frac{\partial (\rho V)}{\partial x} = \nu(x,t)
\label{contin}
\end{equation}
for the vehicle density $\rho(x,t)$. 
The source term $\nu(x,t)\,dx$ is 
the rate of vehicles entering or leaving the freeway at on- or
off-ramp sections of length $dx$.
The differences between the various existing macroscopic
traffic models mainly concern the equation for the average vehicle velocity
$V(x,t)$. The Lighthill-Whitham model \cite{LW,Richards}
and its variants \cite{New,Dag,Dag2,Hil95,Lebacque} assume the
equilibrium relation $V(x,t) = V_{\rm e}(\rho(x,t))$. For the description
of emergent traffic jams and stop-and-go traffic, however, one needs
a dynamic velocity equation. Most proposals can (in their continuous form)
be summarized by the velocity equation
\begin{equation}
 \frac{\partial V}{\partial t}  
 + \underbrace{V \frac{\partial V}{\partial x}}_{\rm Transport\ Term}
 + \underbrace{\frac{1}{\rho} \frac{\partial {P}} 
 {\partial x}}_{\rm Pressure\ Term}
 = \underbrace{\frac{1}{\tau} ( V_{\rm e} - V )}_{\rm Relaxation\ Term} .
\label{geschwin}
\end{equation} 
Their main difference is the specification of the traffic pressure
${P}$, the relaxation time $\tau$, and the dynamic equilibrium velocity
$V_{\rm e}$, which depends on the local vehicle density $\rho$.
Notice that the Lighthill-Whitham model
results in the limit $\tau \rightarrow 0$. 
Payne's \cite{Payne} and Papageorgiou's \cite{Papa} 
model is obtained for ${P}(\rho) = 
[V_0 - V_{\rm e}(\rho)]/(2\tau)$, with the
``free'' or ``desired'' average velocity $V_0 = V_{\rm e}(0)$. For
$d{P}/d\rho = - \rho/[2\tau (\rho + \kappa)] dV_{\rm e}/d\rho$,
one ends up with Cremer's \cite{Cremer} model. 
In the model of Phillips \cite{phillips-kin}, there is
${P} = \rho \theta$, where $\theta$ denotes the velocity variance.
The model of K\"uhne \cite{Kue84,Kue87}, Kerner
and Konh\"auser \cite{kk-93}, and Lee {\em et al.} \cite{Lee} (KKKL model)
results for ${P} = \rho \theta_0 - \eta \partial V/\partial x$, where
$\theta_0$ is a positive constant and $\eta$ some viscosity coefficient.
In comparison with a similar model by Whitham \cite{Whitham},
the contribution $- \eta \partial V/\partial x$ implies an additional 
viscosity term $(\eta/\rho) \,\partial^2 V/\partial x^2$. This is essential
for smoothing shock fronts, which is desireable from empirical and
numerical points of view (but causes some inconsistencies for large
density or velocity gradients \cite{Dag95}, which can be
avoided by a {\em non-local} macroscopic traffic model \cite{num}). 
\par
Many of the above mentioned models have proved their 
value in various applications, but they
have been seriously criticized by Daganzo \cite{Dag95} because of theoretical
inconsistencies. Apart from that, it is
hard to decide which model is the best one. 
Whereas classical approaches focussed on reproducing the empirically 
observed velocity-density relation and the regime of unstable
traffic flow, recent publications pointed out that it is more
important to have traffic models which are able to describe the observed 
spectrum of non-linear phenomena and their characteristic properties
\cite{kk-93,kk-94,kerner-rehb96,bando,GKT,zellauto}. 
We think that it would be desireable to develop
models that reproduce both aspects of empirical data. 
In the following, we will propose such a macroscopic traffic model,
which is theoretically consistent as well \cite{pre,physa}. 

\subsection{The Non-Local, Gas-Kinetic-Based Traffic Model
(GKT model)} \label{Model}

We derived macroscopic traffic equations from a gas-kinetic
(Boltzmann-like) traffic model that was obtained from a simple
microscopic car-following model under the approximation
of quasi-instantaneous braking interactions \cite{book,GKT} 
(which will be circumvented by the micro-macro link suggested in
Sec.~\ref{micro-macro}). Boltzmann-like traffic models for
the spatio-temporal evolution of the so-called phase space density
(= vehicle density $\rho(x,t)$ times velocity distribution $d(v;x,t)$)
\cite{prigogine,paveri-fontana,phillips-kin,nelson,pre,physa,wagner-96,klar-97,Bovy,alex,hoog,kw-98a,kw-98b} 
go back to Prigogine and coworkers (for an overview see Ref.~\cite{prigogine}).
Paveri-Fontana improved Prigogine's model and cured some
of its inconsistencies \cite{paveri-fontana}. Our gas-kinetic
model can be viewed as an extension of Paveri-Fontana's equations to
Enskog-like equations which are also valid in
the regime of moderate and dense traffic, since they take into account
the finite space requirements of vehicles. In particular, the
equations become non-local to account for the fact that drivers
react to the traffic situation ahead of them. This makes the
derivation of the macroscopic traffic equations much more difficult
than for the previous gas-kinetic models, the validity of which was
limited to low vehicle densities. Therefore, our first calculations 
assumed ``vehicular chaos'' (i.e. uncorrelated vehicle velocities)
and a gradient expansion in the vehicle density $\rho(x,t)$, 
average velocity $V(x,t) = \int dv \; v d(v;x,t)$, 
and velocity variance $\theta(x,t) = \int dv \; [v-V(x,t)]^2 d(v;x,t)$
\cite{physa}. In this way, we could derive necessary high-density
corrections of the traffic pressure $P(x,t) = \rho(x,t)\theta(x,t)$
(avoiding that vehicles would otherwise accelerate into jammed
regions), and we could derive a plausible viscosity term, which
had to be introduced in a phenomenological way in previous macroscopic
traffic models. Meanwhile, we managed to carry out the calculations without
a gradient expansion or other approximations 
\cite{parisi,drygran,GKT}, and even velocity
correlations $r$ among successive vehicles 
could be taken into account \cite{mul-HS} (see Sec.~\ref{velcor}).
We have also generalized our model to a multi-class multi-lane
model \cite{book,mul-HS}, which allows to
account for different kinds of
driver-vehicle units and lane changes in an explicit way. However,
the results obtained up to now indicate that an efficient one-lane
model (which averages over the neighboring lanes and the
different driver-vehicle classes) can already capture the macroscopic
vehicle dynamics in a semi-quantitative way \cite{mul-HS,emp-Hel,book}.
Neverthess, in order to reproduce fluctuation effects
like the wide scattering of congested flow-density data 
\cite{Leu88,kerner-rehb96-2},
it is necessary to take the diversity of driver-vehicle types 
into account \cite{GKT-scatter} (see Secs.~\ref{synchro} and \ref{sec:multi}).
\par
In the following, we will discuss the efficient one-lane version
of our non-local, gas-kinetic-based traffic model for the case
of uncorrelated vehicle velocities, but we will come back to 
correlation effects later on. The results of our excessive
calculations can be represented in terms of 
the continuity equation (\ref{contin}) and the velocity equation
(\ref{geschwin}), but with a non-local, {\em dynamical} equilibrium
velocity $V_{\rm e}$ that depends on the vehicle density
and other macroscopic quantities at the actual vehicle location
$x$ and at the advanced ``interaction point'' $x'$: 
\begin{equation}
 V_{\rm e}  = V_0 - \underbrace{\tau [1-p(\rho')]\chi(\rho') \rho' 
 B(\Delta V,S)}_{\rm Braking\ Term} \, .
\label{eqVderiv}
\end{equation}
Apart from other things, this implies that $V_{\rm e}$ 
implicitly depends on the (usually finite) 
variance $\theta$ 
and on the gradients of macroscopic variables like
the density. A prime indicates that the corresponding 
variable is taken at the interaction
point $x'=(x+s)$ rather than at the actual position $x$. For simplicity,
we have assumed the safe distance $s$ to increase linearly with the
average vehicle velocity $V$,
\begin{equation} 
 s = \gamma \left( \frac{1}{\rho_{\rm max}} + TV \right) \, , 
\end{equation}
where $\rho_{\rm max}$ denotes the maximum vehicle density,
$T \approx 1.8$\,s the safe time headway, and $\gamma \approx 1$ an
anticipation factor.
\par
The ``Boltzmann factor''
\begin{equation}  
\label{B}
B(\Delta V,S) =  S \left\{ 
    \Delta V N(\Delta V) 
           + [1+(\Delta V)^2] E(\Delta V) \right\} \, ,
\end{equation}
in which 
\begin{equation}
 N(y) = \frac{\mbox{e}^{-y^2/2}}{\sqrt{2\pi}}
\end{equation}
represents the normal distribution and 
\begin{equation}
 E(y) = \int\limits_{-\infty}^{y} dz \, \frac{\mbox{e}^{-z^2/2}}{\sqrt{2\pi}}
\end{equation}
the Gaussian error function,
describes the dependence of the braking interaction on
the effective dimensionless difference $\Delta V$ between the velocities
at the actual position and the interaction point. The effective velocity
difference is defined by
\begin{equation}
 \Delta V = \frac{V-V'}{\sqrt{S}} \, ,
\end{equation}
where 
\begin{equation}
 S = \theta  + \theta' \, . 
\label{SSS}
\end{equation}
Hence, apart from the velocity difference to the interaction point,
$\Delta V$ takes also into account the velocity variances $\theta$ 
and $\theta'$ at the actual
vehicle position $x$ and the interaction point $x'$, respectively.
In spatially homogeneous traffic, we have $B(0,S)=S/2$. In the limiting case
$\Delta V \gg 0$, where the preceding cars are much slower, we have a much
stronger interaction
$B(\Delta V,S) = (\Delta V)^2 S$. If, in contrast, the preceding cars are
much faster (i.e. $\Delta V \ll 0$), we have
$B(\Delta V,S) \approx 0$. That is, since the distance is increasing, then, 
the vehicle at position $x$ will not brake, 
even if its headway is smaller than the safe distance.
\par
Notice that, for $\gamma=1$, the macroscopic interaction term can be 
easily understood by the underlying microscopic dynamics of the 
gas-kinetic-based traffic model.
If a vehicle at location $x$ with  velocity 
$v$ is faster than one at $x'$ with velocity $w$ 
(i.e. $\Delta v = v-w > 0$), it approaches 
the  car in front within the time 
$\Delta t = \Delta x_{\rm free}/\Delta v$,
where $\Delta x_{\rm free} = 1/(\rho'\chi')$
is the average interaction-free distance per car.
Then, if it cannot overtake immediately, which would happen with
probability $(1-p')$, it abruptly reduces the velocity by
$\Delta v$. The resulting ensemble-averaged deceleration is
\begin{equation}
\label{erwdvdt}
\langle \Delta v/\Delta t \rangle
  = - (1-p') \chi'
  \rho' \int\limits^{\infty}_0 d (\Delta v) \ (\Delta v)^2 d_{_\Delta}(\Delta v) \, .
\end{equation}
If $v$ and $w$ are Gaussian distributed 
with averages $V$, $V'$ and variances $\theta$, 
$\theta'$, respectively, the distribution
function $d_{_\Delta}(\Delta v)$ of the velocity difference $\Delta v$
is Gaussian distributed, with expectation value $(V-V')
= \sqrt{S} \, \Delta V$ and variance $S$.
Evaluating integral (\ref{erwdvdt})
yields $\langle \Delta v/\Delta t \rangle = 
 - (1-p')\chi' \rho' B(\Delta V,S)$, i.e.,
the macroscopic braking term in equation (\ref{eqVderiv}).
\par
Next, we will specify the pressure relation. Similar to gases,
the calculations give
\begin{equation}
 {P} = \rho [\theta + W(\rho)]
\end{equation}
\cite{book,fund}. Again, 
\begin{equation}
 \theta = \langle \theta_i \rangle = \sum_{i=1}^I \frac{\rho_i}{I\rho} \,
 \theta_i
\end{equation}
is the (weighted) average of the lane-specific
velocity variances $\theta_i$, 
where the weights are the relative
densities $\rho_i/(I\rho)$ in the $I$ different lanes $i$.
The monotonically decreasing, density-dependent function 
\begin{equation}
 W = \langle (V_i - V)^2 \rangle 
 = \sum_{i=1}^I \frac{\rho_i}{I\rho} (V_i)^2 
 - \left( \sum_{i=1}^I \frac{\rho_i}{I \rho} V_i \right)^2 
\label{We}
\end{equation}
can be used to correct for the variation of the average velocities 
$V_i$ in the different lanes [see Figure~\ref{suppl}(a)], 
but it is often neglected.
\par
\begin{figure}
\begin{minipage}{1\textwidth}
\includegraphics[width=\textwidth]{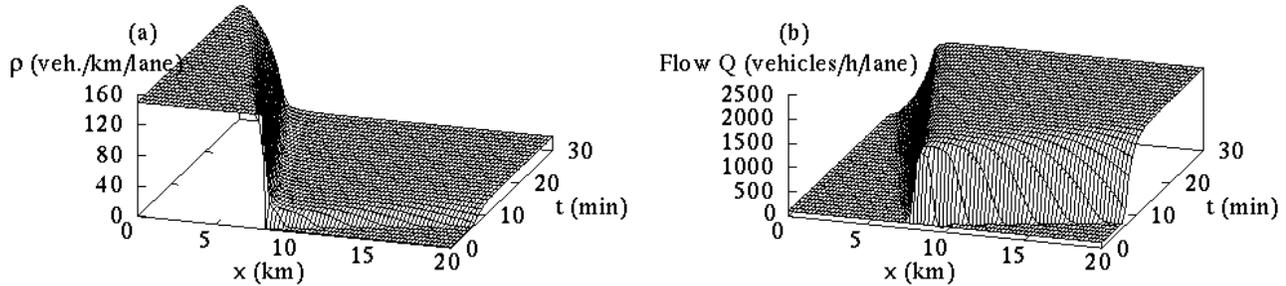}
\end{minipage}
\caption[]{Simulation of a downstream front 
with a jam density of 140 vehicles/km/lane.
Shown is the spatio-temporal development (a) of the density
$\rho(x,t)$, and (b) the flow $Q(x,t) = \rho(x,t)V(x,t)$. 
Free boundary conditions were used on both sides.}
\label{downfront}
\end{figure}

The pressure term describes the kinematic dispersion of
the macroscopic velocity in inhomogeneous traffic as a consequence
of the finite velocity variance. For example, the macroscopic velocity
in front of a small  vehicle cluster will increase {\it even if no
individual vehicle accelerates}, because the faster
cars will leave the cluster behind. 
The kinematic dispersion also leads to a smooth density
profile at the dissolution front between congested
traffic and an empty road, as it occurs 
when a road blockage is removed (see Figure~\ref{downfront} and
Ref.~\cite{num}). 
\begin{center}
{\em Insert Figure \ref{downfront} about here.}
\end{center}
\par
Notice that the total variance 
\begin{equation}
 \Theta = (\theta + W)
\label{Theta}
\end{equation} 
decreases
with density and vanishes at the maximum vehicle density $\rho_{\rm max}$.
Therefore, the density-gradient $dP/d\rho$ of the traffic pressure $P$
may decrease with growing density and increase with decreasing density. 
Hence, according to the
pressure term $(1/\rho)\partial P/\partial x$, it seems that vehicles
would accelerate into denser regions and decelerate when driving into
regions of lighter traffic. This tendency, however, is
overcompensated for by the non-local interaction term
\cite{physa}. Because of its non-locality, the braking term
in relation (\ref{eqVderiv}) strongly increases with the density gradient.
This can be more explicitly seen by applying the gradient
expansions 
$\rho' \approx \rho + s \, \partial \rho/\partial x + \dots$ etc.
\par
\begin{figure}
\begin{center}
  \includegraphics[width=\half]{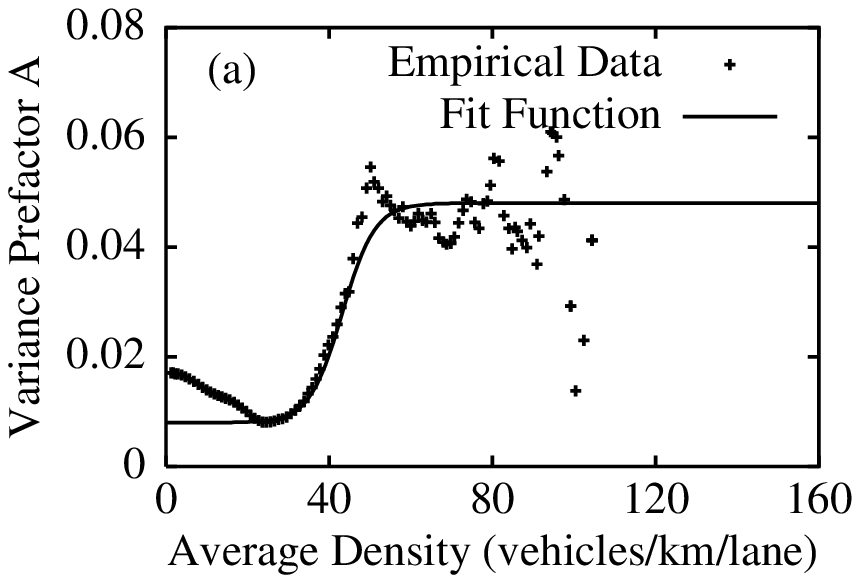}
  \includegraphics[width=\half]{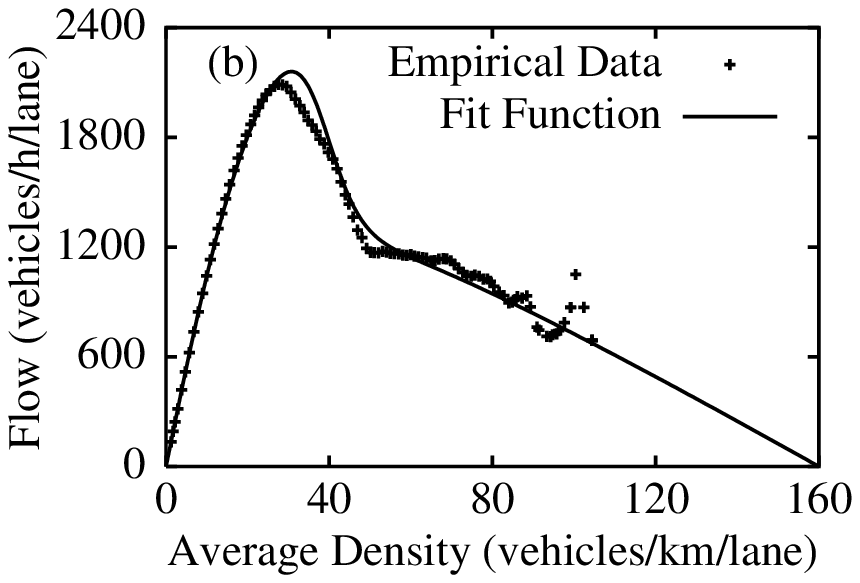}
\end{center}
\caption[]{ \label{fig_AQemp}
Comparison of (a) the density-dependent
relative variance $\Theta$ in units of the squared average velocity,
and (b) the equilibrium flow-density relation
\protect(\ref{Qe}) of our gas-kinetic-based traffic model 
(solid lines) with empirical data (crosses).
The empirical data were obtained from single-vehicle data of the Dutch
motorway A9 (where a speed limit of 120\ km/h applies),
by averaging over one-minute intervals. Note that
both, the variance-density relation and the velocity-density relation
in equilibrium, are fitted by a single, density-dependent function
$A(\rho)$. It turns out that the deviation of the variance-density 
relation from the empirical data at small densities is not of great
importance for the dynamics of the traffic model. It can, however, be
accounted for by the function $W(\rho)$, see formulas (\ref{We})
and (\ref{Theta}).
}
\end{figure}

Next, in order to close the system of equations, one needs to specify 
the dynamics of the velocity variance $\theta$. It turned out that, 
for a description of the presently known properties
of traffic flows, we do not need to describe the variance by
the associated partial differential equation. 
It seems sufficient to describe the variance by the equilibrium relation
\begin{equation}
 \theta = D\tau
\label{Dtau}
\end{equation}
of the dynamical variance equation \cite{emp-Hel}, where $D$ denotes a
velocity diffusion coefficient which is a function of the vehicle
density $\rho$ and the average velocity $V$.
Empirical data suggest that the variance
is a density-dependent fraction $A(\rho)$ of the squared average 
velocity \cite{physa,GKT},
\begin{equation}
 \theta = A(\rho) V^2 \, . 
\label{th1}
\end{equation}
This guarantees that the velocity variance will vanish if the average
velocity goes to zero, as required, but it will be finite otherwise. 
It turns out
that the variance prefactor $A$ is higher in congested traffic than in
free traffic [Fig.~\ref{fig_AQemp}(a)]. 
The empirical data can be approximated by the 
Fermi function 
\begin{equation}
  A(\rho) = A_0 + \Delta A 
 \left[ \tanh \left( \frac{\rho - \rho_{\rm c}}
 {\Delta \rho} \right) + 1 \right] \, ,
\label{th2}
\end{equation}
where $A_0$ and $A_0 + 2\Delta A$ 
are about the variance prefactors 
for free and congested traffic, respectively, $\rho_{\rm c}$ is
roughly of the order
of the critical density for the transition from free to congested traffic,
and $\Delta \rho$ denotes the width of the transition region.
\begin{center}
{\em Insert Figure \ref{fig_AQemp} about here.}
\end{center}
\par
Finally, we will specify the ``effective cross section''
$(1-p)\chi = (1-p)/p$, assuming that the average time headway of vehicles
corresponds to the safe time headway $T$,
if the vehicle density is high (i.e. the
freeway space is almost used up completely by the vehicular space
requirements). This specification means that, in dense and homogeneous traffic
situations with $V' = V$ and $\theta' = \theta$, we should have
$s = 1/\rho = [1/\rho_{\rm max} + T V_{\rm e}(\rho)]$, where we insert
the equilibrium solution (\ref{Ve}). This gives us the relation
\begin{equation}
 [1-p(\rho)]\chi(\rho) = \frac{V_0 \rho T^2}{\tau A(\rho_{\rm max}) 
 (1-\rho/\rho_{\rm max})^2} \, ,
\label{bleibt}
\end{equation}
which makes also sense in the low-density limit $\rho \to 0$, since it
implies $\chi \to 1$ and $p\to 1$. 

\subsection{Properties of the Non-Local, Gas-Kinetic-Based 
Traffic Model} \label{properties}

\begin{figure}[htbp]
\begin{center}
\begin{minipage}{1\textwidth}
\includegraphics[width=\textwidth]{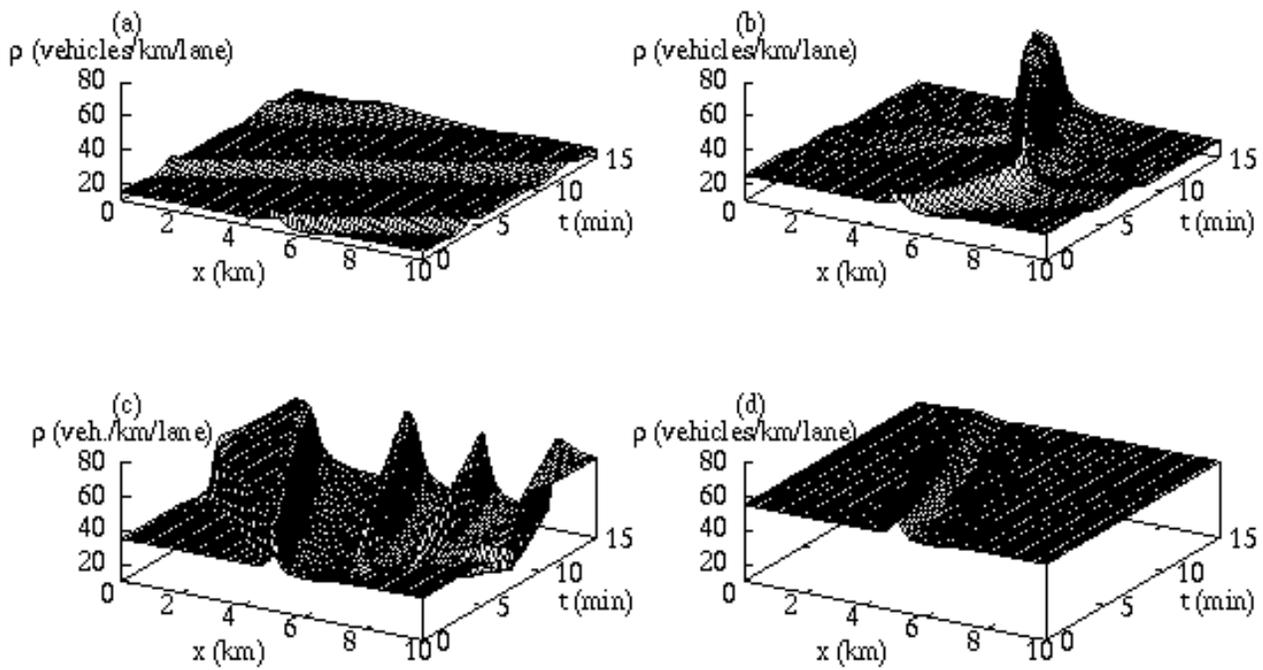}
\end{minipage}
\end{center}
\caption[]{ \label{stop_go} Spatio-temporal evolution of the
traffic density $\rho(x,t)$ on a uni-directional ring of circumference 10 km,
starting with homogeneous traffic to which
a localized initial perturbation of amplitude $10$ vehicles/km/lane
is added. 
(a) Free and stable traffic at an average density of 15 vehicles/km/lane,
(b) metastable traffic for 25~vehicles/km/lane,
(c) unstable traffic for 35~vehicles/km/lane,
(d) stable congested traffic for 55~vehicles/km/lane.
The chosen model parameters are displayed in
Table~\protect\ref{tab_par}.
}
\end{figure}

Summarizing the above results, our gas-kinetic-based traffic model can
be written in the general form (\ref{contin}), (\ref{geschwin}) of
macroscopic traffic models, but with a non-local relaxation term.
Since the equations are structurally related to,
for example, the KKKL model, we find
many similar non-linear phenomena \cite{GKT}. 
This includes the sequence of stable, linearly unstable,
and metastable regimes \cite{kk-94,kerner-dipole} 
the local breakdown effect \cite{kerner-ramp},
the local cluster effect \cite{kk-94}, 
and, at sufficiently large densities, 
the formation of so-called dipole layers \cite{kerner-dipole}.
The response of equilibrium traffic to
localized disturbances 
is similar to the Kerner-Konh{\"a}user model \cite{kerner-ramp}.
More specifically, for densities below a
certain critical density $\rho_{\rm c1}$ and above some density
$\rho_{\rm c4}$, homogeneous traffic 
is stable with respect to localized perturbations
[Figure~\ref{stop_go}(a)+(d)], 
and for a range $\rho_{\rm c2} < \rho < \rho_{\rm c3}$ 
of intermediate densities, it is 
linearly unstable, giving rise to cascades of traffic jams 
[``stop-and-go traffic'', cf. Figure~\ref{stop_go}(c)]. For the two density regimes 
$\rho_{\rm c1} \le \rho \le \rho_{\rm c2}$ and 
$\rho_{\rm c3} \le \rho \le \rho_{\rm c4}$ between the stable and
the linearly unstable regions, traffic is metastable, i.e., it behaves
unstable with respect to perturbations exceeding a 
certain critical amplitude $\Delta \rho_{\rm cr}(\rho)$
[Figure~\ref{stop_go}(b)], 
but otherwise stable. For 
the self-organized density $\rho_{\rm jam}$ inside traffic jams 
we find a typical value $\rho_{\rm jam} \ge \rho_{\rm c4}$ 
\cite{GKT}. Furthermore, there exists a range 
$\rho_{\rm cv} < \rho < \rho_{\rm c3}$ with
$\rho_{\rm cv} > \rho_{\rm c2}$, where traffic is linearly
unstable, but convectively stable, i.e., all perturbations grow, but
they are eventually convected away in upstream direction
\cite{Cross}.
\begin{center}
{\em Insert Figure \ref{stop_go} about here.}
\end{center}
\par
\begin{figure}
\begin{center}
  \includegraphics[width=\half]{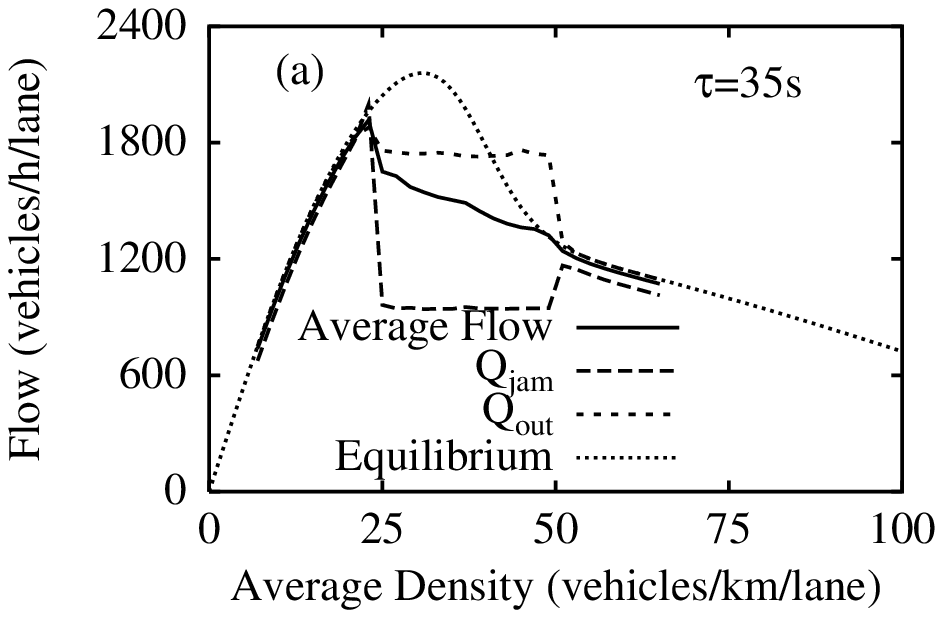}  
  \includegraphics[width=\half]{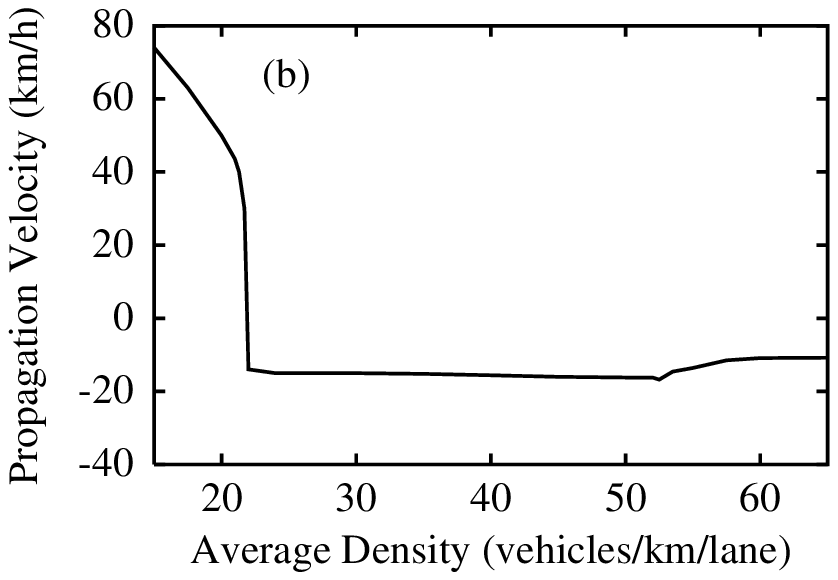}  
\end{center}

\caption[]{\label{ampli}(a) Characteristic flows resulting in 
fully developed stop-and-go traffic on a circular road, as a function 
of the spatially averaged density. 
Depicted are the flows $Q_{\rm jam}$ in the jammed regions (--~--),
the outflows $Q_{\rm out}$ from jams (-~-~-),
and the average flows (---). For comparison, the
equilibrium flow $Q_{\rm e}(\rho) = \rho V_{\rm e}(\rho)$ 
with $V_{\rm e}(\rho)$ from
equation \protect(\ref{Ve}) is also shown ($\cdots$).
Notice that, in the unstable range, the average {\it dynamic} flow is
lower than the equilibrium flow. 
(b) Propagation velocity of density clusters as a function of average density.
We started our simulations with an initial perturbation of amplitude
10\ vehicles/km/lane and measured the group velocity 
after a sufficiently long transient time. 
The propagation velocity at low densities is
positive, but slower than the average vehicle velocity. In the 
instability region, the negative propagation 
velocity of fully developed traffic jams (i.e. their dissolution velocity) 
is independent of
the initial density, and its magnitude is in agreement 
with empirical data. Notice that the dissolution velocity $C$ of jam
fronts corresponds to the slope of the average flow depicted in (a).
}

\end{figure}

In addition, we obtain that, in the unstable traffic regime, the resulting 
flow-density relation differs from the equilibrium one 
(lying below the latter). We also find that the outflow $Q_{\rm out}$
from traffic jams is independent of the initial conditions and the
spatially averaged density [Figure~\ref{ampli}(a)]. 
Furthermore, the dissolution velocity $C$ of
traffic jams varies only a little with density [Figure~\ref{ampli}(b)]. 
\begin{center}
{\em Insert Figure \ref{ampli} about here.}
\end{center}
\par
\begin{figure}
\begin{minipage}{1\textwidth}
\includegraphics[width=\textwidth]{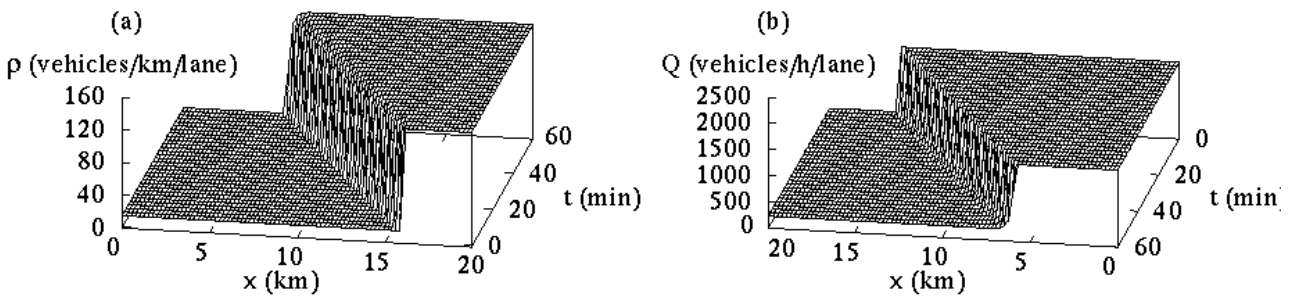}
\end{minipage}
\caption[]{
Simulation of an upstream front with initial densities
of $\rho_{1}=15$ vehicles/km/lane
and $\rho_{2}=140$ vehicles/km/lane.
Shown is the evolution of (a) the density $\rho(x,t)$, and (b) the 
flow $Q(x,t) = \rho(x,t)V(x,t)$.
Note that, in contrast to other macroscopic traffic models \cite{Dag95},
the simulations do {\em not} produce
densities above $\rho_{\rm max}$ or negative flows.
In (b), the direction of the space and time axes is reversed for
illustrative reasons.
}
\label{upfront}
\end{figure}

The main difference with respect to other macroscopic traffic models is
the non-local character of the braking term. Nevertheless, our model 
can still be rewritten in the form of flux equations, with a non-local 
source term \cite{parisi,drygran}. For this reason, we can
apply various standard methods of numerical integration. It turns
out that the non-local term has similar smoothing
properties like a viscosity term. However, its effect is 
anisotropic due to the anticipation behavior in forward direction
reflecting that vehicles mainly react on density or velocity gradients 
ahead of them, not behind them (compare Fig.~\ref{upfront}
with Fig.~\ref{downfront}).
Compared to a viscosity term, the non-locality does not change the 
hyperbolic character of the partial differential equations to a parabolic 
one. Therefore, it has 
favourable properties with respect to numerical stability and integration
speed \cite{num}. As a consequence, our model
allows a robust real-time simulation of freeway stretches up to several
thousand kilometers on a usual PC. 
\begin{center}
{\em Insert Figure \ref{upfront} about here.}
\end{center}

\subsection{Model Calibration and Validation}

It turns out that our model can be easily calibrated to the static and
dynamic properties of traffic flow data by a systematic
procedure. In spatially homogeneous traffic, the average velocity relaxes to
\begin{equation}
\label{Ve}
V_{\rm e}(\rho) = \frac{\tilde{V}^2}{2V_0}
       \left( - 1 + \sqrt{1 + \frac{4 V_0^2}{\tilde{V}^2}}
       \right)
\end{equation}
with 
\begin{equation}
\label{tilV}
\tilde{V}(\rho)  
 =\sqrt{\frac{V_0}{\tau  [1-p(\rho)]
  \chi(\rho) \rho A(\rho)} } 
 = \frac{1}{T} \left(\frac{1}{\rho}-\frac{1}{\rho_{\rm max}}
 \right) \sqrt{\frac{A(\rho_{\rm max})}{A(\rho)} } \, .
\end{equation}
This also determines the equilibrium traffic flow per lane by
\begin{equation}
\label{Qe}
  Q_{\rm e}(\rho) = \rho V_{\rm e}(\rho) \, ,
\end{equation}
which is depicted in Fig.~\ref{fig_AQemp}(b). 
In the limit of small densities we find $V_{\rm e}(\rho) \approx V_0$,
while in the limit of high densities (i.e. for $(1-\rho/\rho_{\rm max}) \ll 1$ 
or $V_{\rm e} \ll V_0$), we have 
$V_{\rm e}(\rho) \approx \tilde{V}(\rho)$, i.e.
$V_{\rm e}(\rho) \approx (1/\rho - 1/\rho_{\rm max})/T$.
Apparently, the equilibrium flow-density relation
is only affected by the model parameters $V_0$, $T$, and $\rho_{\rm max}$.
The desired velocity $V_0$ can be determined by fitting 
flow-density data at low densities by a straight line $\rho V_0$,
while the safe time headway $T$ and the maximum density $\rho_{\rm max}$
are obtained by fitting the data at high densities by a straight line
$(1-\rho/\rho_{\rm max})/T$, which 
crosses the abszissa at $\rho_{\rm max}$ with a slope of
$-1/(\rho_{\rm max}T)$.
\par
\begin{figure}
\begin{center}
   \includegraphics[width=\half]{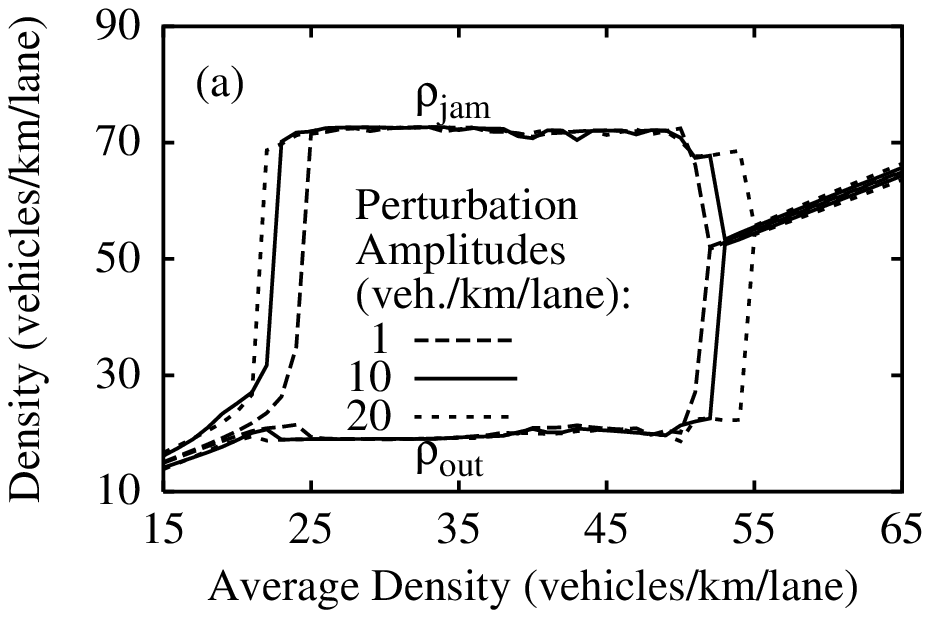}  
   \includegraphics[width=\half]{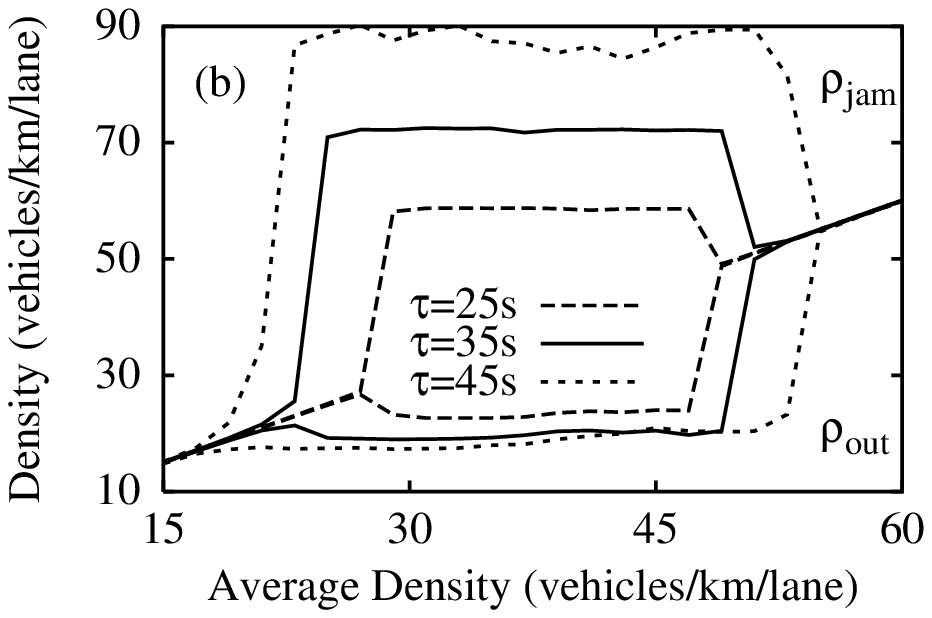} 
\end{center}

\caption[]{ \label{phasediag} Stability diagram 
for a localized perturbation of homogeneous traffic 
on a ring of circumference 10 km.
Both diagrams show
the developing maximum and minimum densities $\rho_{\rm jam}$
and  $\rho_{\rm out}$ as a function of the spatially averaged density,
measured after a dynamical equilibrium was reached. 
The unstable traffic regime corresponds to the density range where 
the jam amplitude
$(\rho_{\rm jam} - \rho_{\rm out})$ is large (rectangle-like
shaped regions).
Diagram (a) shows the dependence of the stability diagram on the 
perturbation amplitude $\Delta \rho$.
One can clearly see two density ranges  
$[\rho_{\rm c1},\rho_{\rm c2}]$ with
$\rho_{\rm c1} =$\, 21 
vehicles/km/lane and $\rho_{\rm c2} = $\, 24 
vehicles/km/lane, and
$[\rho_{\rm c3},\rho_{\rm c4}]$ with
$\rho_{\rm c3} =$\, 51 
vehicles/km/lane and $\rho_{\rm c4} = $\, 55 
vehicles/km/lane, where  traffic is non-linearly stable, i.e.,
stable for small perturbations, but unstable for large
perturbations. In the range
$[\rho_{\rm c2},\rho_{\rm c3}]$, 
homogeneous traffic is unstable for arbitrary
perturbation amplitudes.
Diagram (b) shows the stability diagram for various relaxation times
$\tau$ and a perturbation
amplitude of $\Delta\rho = 1$\,vehicle/km/lane.
}
\end{figure}

The model parameters $\tau$ and
$\gamma$ influence the stability behavior.
Figure~\ref{phasediag}
shows that an increased value of
$\tau$ leads both to an increased range of {\em instability}, and
to increased amplitudes 
$(\rho_{\rm jam} - \rho_{\rm out})$ of traffic jams. 
Further simulations showed that higher values of
$\gamma$ tend to increase the {\em stability} of 
traffic. This is plausible, since  $\gamma$ reflects the 
anticipation of  future velocity changes. $\gamma$ and $\tau$
also determine the shape and width of the downstream and upstream fronts 
connecting free and congested states
(see Figs.~\ref{downfront} and \ref{upfront}).
Since $\tau$ and $\gamma$ weakly influence the outflow 
$Q_{\rm out}$ from traffic jams,
the calibration of $T$, $\tau$, and $\gamma$ is repeated recursively until
an optimal fit of the model to the empirical data is achieved.
\begin{center}
{\em Insert Figure \ref{phasediag} about here.}
\end{center}
\par
Because of their intuitive meaning,
the plausible range of values for the model parameters
is rather restricted. $V_0$ is given by the average free speed.
$\rho_{\rm max}$ must be consistent with the
average length of vehicles plus 
a reasonable bumper-to-bumper distance of about 
1.5\,m in standing traffic.
$T$ should be compatible 
with the average time headway kept in homogeneous congested traffic.
Reasonable (average) time headways are in the range 
$T \in $ [1.0 s, 2.5 s]. Initial accelerations $a_{\rm max} = V_0/\tau$
are typically in the range $a_{\rm max} \in $ [1 m/s$^2$, 4 m/s$^2$],
corresponding to $\tau \in$ [10\,s, 40\,s]. (Notice that, for
$V_0$ = 158 km/h, $\tau$ would have the meaning of the average acceleration
time from 0 to 100 km/h.) Therefore, a relaxation
time $\tau \approx 35$\,s is reasonable for freeway traffic, whereas
$\tau$ is smaller for city traffic. 
Finally, the minimum anticipation of traffic is to the car in front,
implying $\gamma \ge 1$.
\par
\begin{figure}
\begin{center}
   \includegraphics[width=\half]{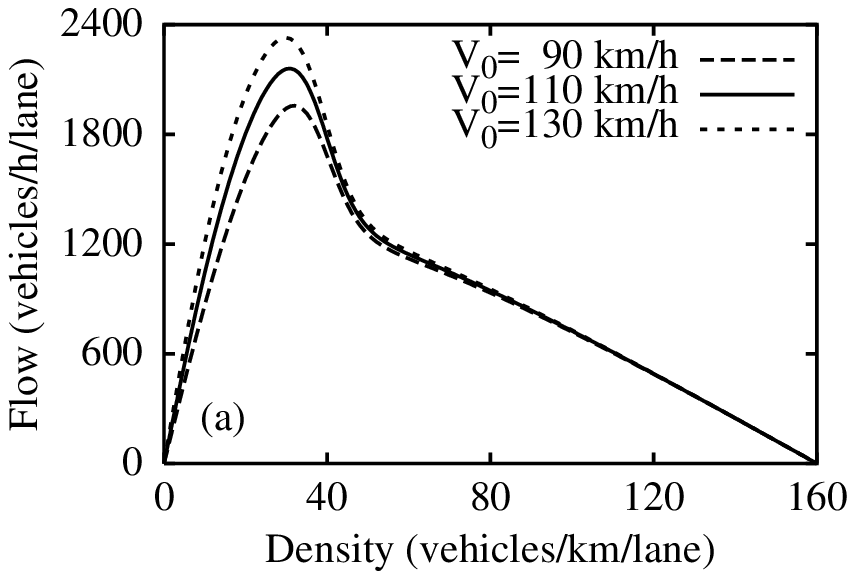} 
   \includegraphics[width=\half]{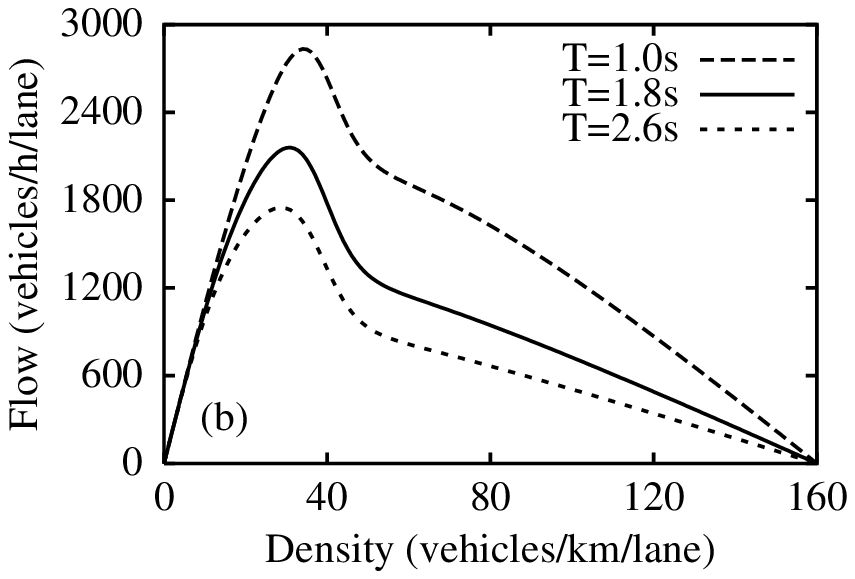} 
\end{center}
\caption[]{ \label{fundamentals} Equilibrium flow-density relations of
the non-local, gas-kinetic-based traffic model.
The diagrams (a) and (b) show the variation with
the model parameters $V_0$ and $T$, respectively. 
Notice that a reasonable speed limit almost does not affect the flow in the
congested regime, while a variation of the following time $t$ can
account for large variations in the flow in the congested density
regime. 
In each diagram, the solid lines correspond to the
standard parameter set displayed in Table 
\protect\ref{tab_par}.}
\end{figure}

The above procedure of parameter calibration
has been applied to single-vehicle data of the Dutch freeway
A9, leading to the parameter set shown in Table~\ref{tab_par}.
It turns out that all optimized parameters have realistic values. 
In particular, this holds for $\tau$. 
The typical values for the safe time headway of $T=1.8$\,s
is consistent with the rule ``distance (in m) should not be 
less than half the velocity
(in km/h)'' suggested by German road authorities.
For other data, however, one often finds
that a somewhat smaller time headway yields a better fit.
\begin{center}
{\em Insert Table \ref{tab_par} about here.}
\end{center}
Since the model parameters are meaningful, it is simple
to model changes of the traffic dynamics caused by external effects
like environmental influences (cf. Figure~\ref{fundamentals}).
For example, a speed limit would be considered by decreasing $V_0$.
Bad weather conditions leading to more defensive driving 
would be characterized by an increased time headway $T$ and a lower
value of $V_0$ (plus a reduction of $\gamma$, if there is heavy fog).
In rush-hour traffic, it is plausible to assume a higher percentage
of experienced drivers than in holiday traffic, which would correspond to
a higher $\gamma$.
Effects like a varying distribution of vehicle types
can be modelled as well.
For example, a higher proportion of lorries (long vehicles)
would lead to a decrease of
$V_0$ and $\rho_{\rm max}$,
but also to an increased value of  $\tau$.
\begin{center}
{\em Insert Figure \ref{fundamentals} about here.}
\end{center}

\subsection{Treatment of Boundaries in Open Systems}

Notice that the left-hand sides of the continuity equation (\ref{contin})
and the velocity equation (\ref{geschwin}) 
constitute a hyperbolic set of partial
differential equations. Hence, the direction of information 
flow due to convective and dispersive
processes is given by the characteristic lines \cite{ames}.
Specifically, the velocities of information flow are given by
\begin{equation}
\lambda_{1,2}=V-\frac{1}{2 \rho}\frac{\partial P}{\partial V}
\pm \sqrt{
\frac{1}{4 \rho ^2}\left(\frac{\partial P}{\partial V}\right)^2
+\frac{\partial P}{\partial \rho}} \quad {\rm with}\quad
P(\rho,V)=\rho\theta(\rho,V) \, . 
\end{equation}
For all physically reasonable parameter sets they are always
positive. Information flow
{\em against} the direction of motion of the simulated vehicles is 
taken into account by the
non-local interaction term.
The consequences are: (i) There is always a transport of information in
both directions, upstream and
downstream. (ii) The most suitable 
differencing scheme for the numerical integration
is the upwind scheme \cite{numrep}.
\par
By choosing the upwind scheme, the downstream boundary conditions are 
only relevant for the non-local interaction term. 
This is in accordance with the boundary conditions for microscopic traffic
simulations, where the first car at the downstream boundary always 
needs a ``phantom'' leading car
for its update process. In particular, the microscopic 
downstream boundary conditions determine only
the {\em acceleration} of this first car, 
but do not determine the {\em flow} of exiting vehicles.
At the same time, there always has to be a defined inflow of 
new vehicles at the upstream boundary.
The same is true for the simulation of non-local macroscopic models 
with the upwind scheme.
Therefore, it is no problem to formulate downstream
boundary conditions which are always satisfied. 
\par
At the upstream boundary,
unphysical situations can emerge when a backwards moving jam 
approaches the boundary, but the assumed boundary (in)flow
is higher than the flow inside the traffic jam.
Then, the equations are no longer well-posed in a mathematical sense.
Such situations may occur, whenever there is a slight 
offset in the arrival times of simulated and real jams at the boundary.
This means the boundary conditions have to be implemented in a way that,
on the one hand, copes with unexpected or offset density waves
and, on the other hand, returns to the exact solution as soon as possible. 
This can be done by an algorithm that implements 
homogeneous von Neumann boundary conditions 
instead of Dirichlet boundary conditions 
for density and flow, 
if (i) there is congested traffic in the vicinity
of the boundary, and (ii) 
the imposed inflow is larger than the traffic flow of the congested
state \cite{num}. Condition (i) means that, near the boundary,
the local group velocity
$v_{\rm g} = \partial (\rho V_e(\rho)) / \partial \rho$
for instantaneous velocity relaxation is negative, or
\begin{equation}
\rho(\delta,t) > \rho_{\rm m}
\end{equation}
(with some small $\delta > 0$ and $\rho_{\rm m}$ being the density for
maximum equilibrium flow). Condition (ii)
can be expressed by requiring
\begin{equation}
Q(0,t) > Q(\delta,t).
\end{equation}
\par
\begin{figure}[h]
\begin{center}
\includegraphics[width=\textwidth]{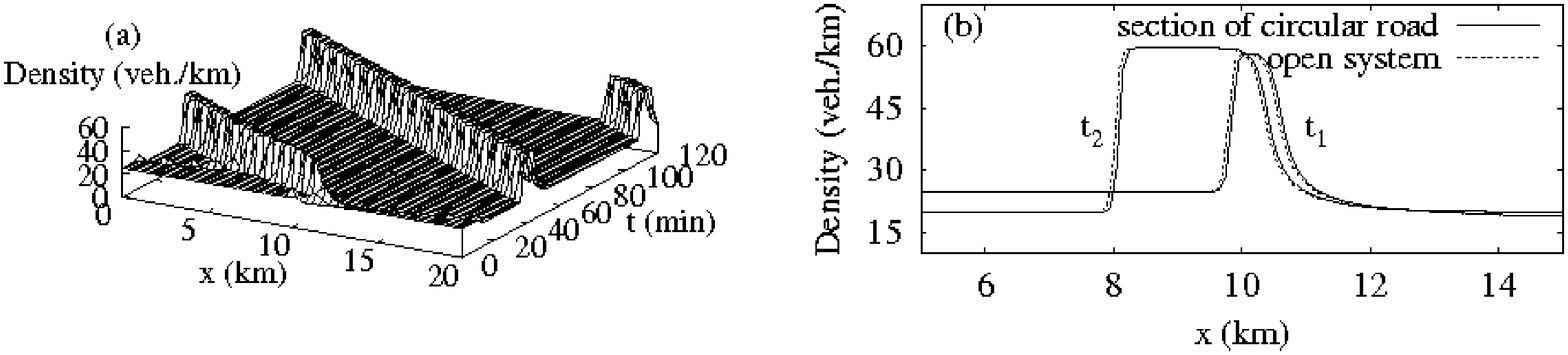}
\caption[]{\protect (a) Circular reference system simulated 
in order to obtain boundary conditions for the simulation of a
comparable open system.\label{fig:scen}
(b) Comparison of density profiles at two times $t_1=8$\,min 
and $t_2=80$\,min,
obtained in the cirular reference system (solid) 
and in an open system which was simulated with boundary 
conditions taken from the reference scenario (dashed).
In order to make the curves distinguishable, 
one of the profiles was shifted by 100\,m.\label{fig:comp}}
\end{center}
\end{figure}

This algorithm has been tested as follows: First, we simulated a
circular road of 20 kilometers length.
The simulated scenario contains a local 
dipole-like perturbation in the initial
conditions which develops to a backwards 
moving cluster [see Fig.~\ref{fig:scen}(a)]. Then, we used the time-dependent 
results at the locations $x=5$\,km and $x=15$\,km as Dirichlet boundary
conditions for the simulation of a new, 10 kilometer long
{\em open} system. Figure \ref{fig:scen}(b) shows a 
comparison of the density profiles of the two systems
at two times $t_1$ and $t_2$. 
At $t_1=8$\,min, the cluster is not yet fully developed, and at
$t_2=80$\,min, it has propagated around the circle once.
The profiles belonging to the same time points agree so well,
that they would lay one upon each other. For this reason, we have
depicted them with a slight artificial offset. After this 
hard and successful test, we are prepated to simulate the
traffic phenomena observed on real freeways, which are open systems.
\begin{center}
{\em Insert Figure \ref{fig:scen} about here.}
\end{center}

\subsection{Simulation of ``Synchronized'' Congested Traffic}\label{synchro}

Recently, Kerner and Rehborn \cite{kerner-sync} 
have presented experimental data
indicating a first-order transition to ``synchronized'' congested
traffic. Traffic data from several freeways 
in Germany \cite{kerner-rehb96-2,kerner-sync}, 
the Netherlands \cite{book,emp-Hel,GKT-scatter},
and the United States \cite{OCT,Daganzo} indicate 
that synchronized traffic is the most common form of congested traffic.
Synchronized traffic typically occurs at on-ramps 
when vehicles are added to already busy ``freeways'' and
has the following properties:
(i) The dynamics of the average velocities on all lanes is highly correlated
(``synchronized'').
(ii) The corresponding points in the flow-density plane are 
widely scattered.
(iii) Synchronized traffic
is characterized by a low average velocity, but, in contrast to
traffic jams, the associated traffic flow is rather high.
(iv) The transition to synchronized traffic is often caused by a localized
and short perturbation of traffic flow that 
starts downstream of the 
on-ramp and propagates upstream with a velocity of about $-10$\,km/h.
(v) As soon as the perturbation passes the
on-ramp, it triggers synchronized traffic 
which spreads upstream in the course of time.
(vi)  Downstream, synchronized traffic eventually relaxes to free traffic.
(vii) Synchronized traffic often persists over several hours. 
(viii) The transition from synchronized traffic 
to free traffic occurs at a lower density and higher average 
velocity than the inverse transition {\em (hysteresis effect).}  
\par
\begin{figure}
\unitlength=1.0cm
\begin{center}
\includegraphics[width=0.85\textwidth]{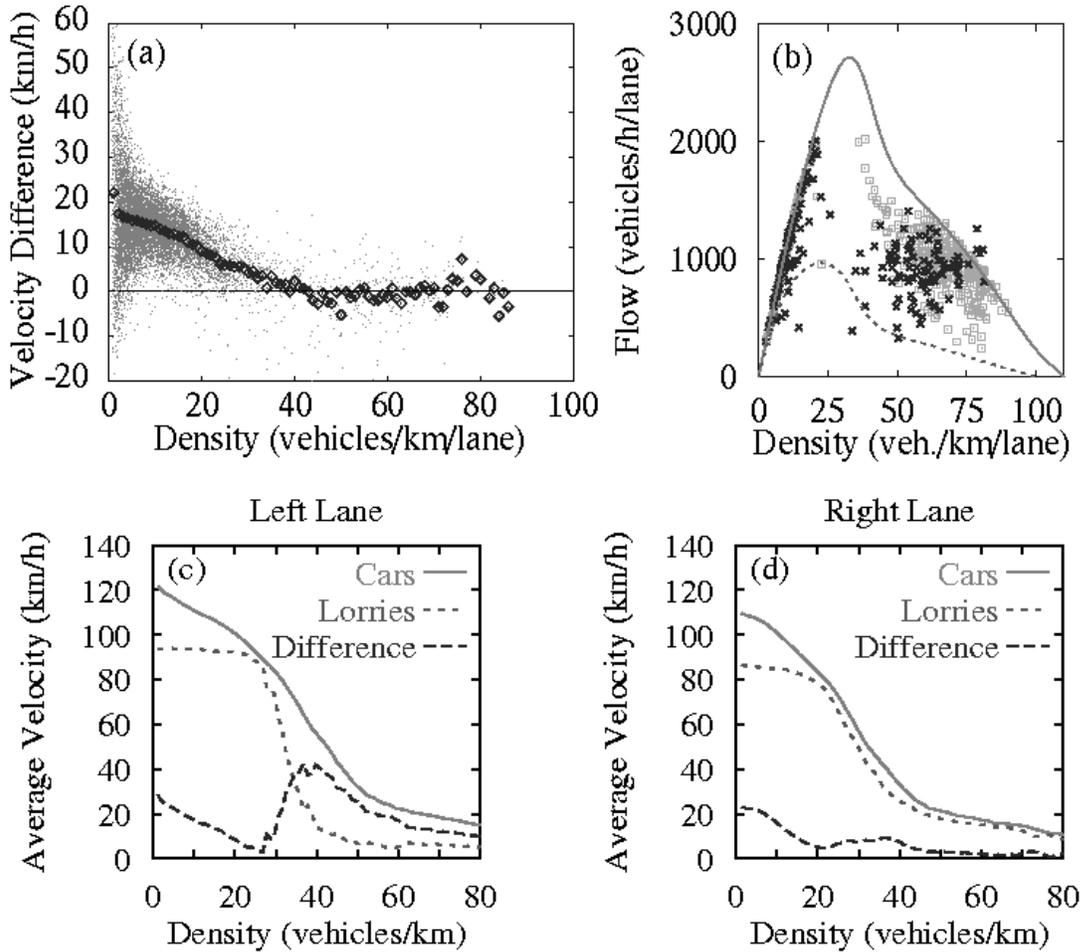} 
\end{center}
\vspace*{-5mm}
\caption[]{(a) Difference between the average velocity $V_2(\rho)$ in the
left lane and $V_1(\rho)$ in the right lane as a function of the
lane-averaged vehicle density $\rho$.  
Points correspond to 1-minute data from November 2, 1994, which were
determined from single-vehicle data of the Dutch 
freeway A9. We find no gap in the data cloud, so that there is no
support for the hypothesis of a phase transition to synchronization
among lanes. The same conclusion can be drawn from the 
averages of the 1-minute data belonging to the same density, which are 
represented by diamonds. We find a linear decrease without discontinuities
up to a density of 40 vehicles per kilometer and lane, 
above which the velocity in both lanes is the same.\\ 
(b) 
The displayed points in flow-density space correspond
to lane-averaged one-minute data 
(dark crosses) and related simulation results (grey boxes)
at the location of an on-ramp (after \cite{GKT-scatter}).
The simulations manage to reproduce both, the quasi-linear 
flow-density relation at small densities and the scattering 
of the data over a two-dimensional region in so-called ``synchronized''
congested traffic at medium and high densities. 
The simulations are based on the assumption
that, in an effective model of mixed traffic, 
the time-dependent parameters $X(t)$ are
given by a linear interpolation $X(t) = \{p_{\rm lo}(t) X_{\rm lo}
     + [1- p_{\rm lo}(t)] X_{\rm car}\}$
between the parameters
$X_{\rm car}$ of cars and $X_{\rm lo}$ of lorries, 
where $p_{\rm lo}(t)$ denotes the 
time-dependent fraction of long vehicles passing the cross section. 
By lines, we have displayed
the assumed equilibrium flow-density relations
for traffic consisting of 100\% cars (---),
and of 100\% long vehicles (-~-~-). 
(c), (d) Average velocities of cars and lorries (long vehicles)
in the left and the right lane as a function of density. The
difference of these empirical curves shows a minimum around 25
vehicles per kilometer, where cars are almost as slow as lorries
(after \cite{helb-nat}). However,
at higher densities, cars are faster again, which shows that there
must be overtaking maneuvers at medium and high densities, which does
not support the hypothesis of synchronization among lanes.\label{suppl}}
\end{figure}

We believe that property (i), i.e., the synchronization 
requires a strong coupling among lanes. When the street becomes
crowded, the drivers are trying to avoid obstructions by changing
lanes. They fill developing gaps in the other lane until
it becomes as slow as the own lane [Fig.~\ref{suppl}(a)],
thereby balancing each upcoming disequilibrium among lanes.
While restoring the balance is related with a considerable number of
lane changes \cite{mul-HS}, the lane changing rate becomes 
rather low as soon as the synchronization among lanes is reached.
This idea can be indirectly supported by empirical traffic data
which show that, even in congested traffic, there can be a significant
difference in the velocities of cars and lorries [Fig.~\ref{suppl}(c), 
(d)], which can be only attributed to lane changing and overtaking.
\begin{center}
{\em Insert Figure \ref{suppl} about here.}
\end{center}
\par
Property (ii) seems to be
the result of the diversity of vehicles \cite{GKT-scatter},
as demonstrated in Figure~\ref{suppl}(b). The other
characteristic properties of ``synchronized'' traffic 
can be reproduced by the
effective one-lane version of the 
non-local, gas-kinetic-based traffic model described above. 
This applies to homogeneous forms of ``synchronized'' traffic as well
as to non-stationary forms (see oscillating congested traffic, OCT,
discussed below). The essential point is that,
along on-ramps (or off-ramps), there is a source term in the
continuity equation (\ref{contin}). With 
\begin{equation}
  \nu = \frac{Q_{\rm rmp}}{nL} \, ,
\end{equation} 
we assume that this source term
is given by the actually observed inflow $Q_{\rm rmp}>0$ 
from (or outflow $Q_{\rm rmp}<0$ 
to) the ramp, divided by
the merging length $L$ and by the number $n$ of lanes.
The inflow has an upper limit that depends on the
downstream flow on the main road \cite{infl}. 
If one likes to model the tendency that some on-ramp vehicles
enter as soon as possible regardless of the length of the ramp, 
one replaces $L$ by an effective ramp
length $L^*<L$. Moreover, if the average velocity
$V_{\rm rmp}$ of entering or leaving vehicles on the ramp is different
from the average velocity $V$ on the main road, this gives an additional
contribution
\begin{equation}
 \frac{\nu}{\rho} (V_{\rm rmp} - V)
\end{equation}
to the right-hand side of the 
velocity equation (\ref{geschwin}), corresponding to a
source term $\nu V_{\rm rmp}$ in the related differential equation
for the traffic flow $Q = \rho V$ on the freeway \cite{book,ML}.
However, usually one assumes $V_{\rm rmp} \approx V$.

\begin{figure}
\begin{center}
   \includegraphics[width=0.9\textwidth]{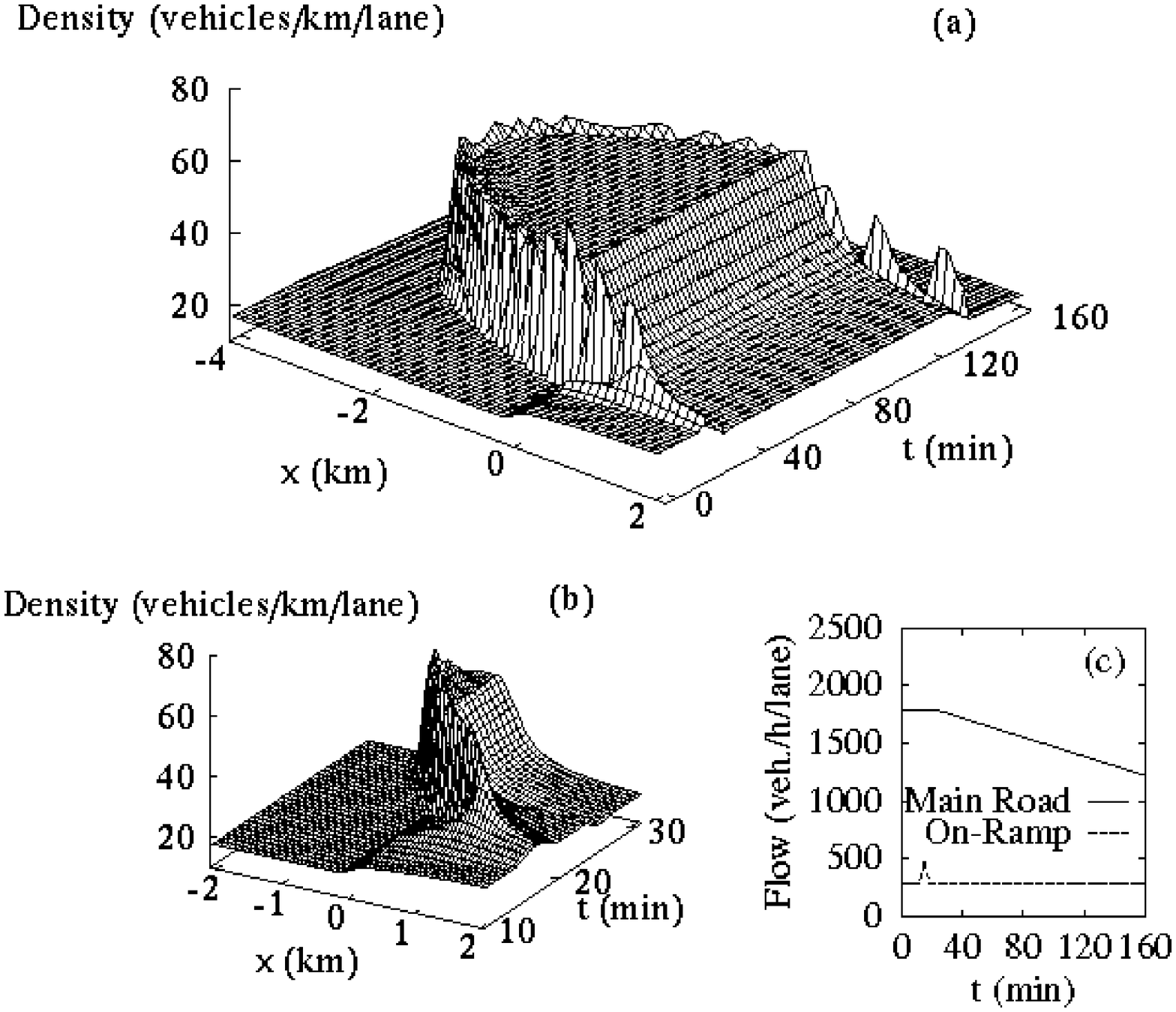} 
\end{center}
\caption[]{Spatio-temporal evolution of the lane-averaged density 
after a small peak of inflow from the on-ramp (after \cite{HT-sync}). 
The on-ramp merges with
the main road at $x=0$\,km with a merging length of $L=300$\,m.
Traffic flows from left to right.
In (a), the parabolically shaped region of high density
corresponds to ``synchronized'' congested traffic.
Plot (b) shows the formation of this state in more detail.
The time-dependent inflows $Q_{\rm main}$ at the upstream boundary 
and $Q_{\rm rmp}/n$ at the on-ramp are displayed in (c).\label{syncfig}} 
\end{figure}

Our simulations reproduced 
the observations (iii) to (viii) very realistically. 
For $V_{\rm rmp} = V$, the parameter values $V_0=128$\,km/h,  
$\rho_{\rm max}=160$\,vehicles/km, 
$T=1.6$\,s,
$\tau=31$\,s, 
$\gamma=1.0$, and the observed 
boundary flows, [Figure~\ref{syncfig}(c)], 
the simulated velocities and flows are, apart from fluctuations,
in good agreement with the empirical curves presented 
by Kerner and Rehborn (compare Ref.~\cite{HT-sync} 
with Ref.~\cite{kerner-sync}). 
\begin{center}
{\em Insert Figure \ref{syncfig} about here.}
\end{center}
\par
Our results suggest the following interpretation of 
the phase transition to ``synchronized'' congested traffic:
Initially, the homogeneous flow $Q_{\rm main}$ upstream of an on-ramp 
is stable, while the higher downstream flow
$Q_{\rm down} = Q_{\rm main} + Q_{\rm rmp}/n$ is metastable. 
Without any disturbance of the main or ramp flow,
free traffic flow will continue. 
However, a (positive or even a negative) perturbation in the
on-ramp or the main flow will eventually
cause a breakdown of velocity close to the on-ramp.
Speaking in the terminology of Daganzo {\em et al.} \cite{Daganzo}, 
the perturbation will 
eventually ``activate the bottleneck'' associated with the on-ramp. 
In the following, we will explain and quantify the mechanism
of this transition. 

According to Figure~\ref{syncfig}, the perturbation
triggers a stop-and-go wave, which travels downstream as long as it is 
small and upstream as it becomes larger, as is 
known from ``localized clusters'' \cite{kerner-ramp}. 
Now, assume the downstream front of the cluster would
pass the on-ramp. Then, since $Q_{\rm main}$ [Figure~\ref{syncfig}(c)] is lower
than the characteristic outflow $Q_{\rm out}$ 
from a jam, the cluster would eventually vanish. 
However, during its lifetime, the cluster would continue to emit
the flow $Q_{\rm out}$, leading downstream of the ramp to a 
flow $Q_{\rm out}+Q_{\rm rmp}/n > Q_{\rm max}$.
As a consequence, 
as soon as the perturbation reaches the on-ramp, 
it induces congested traffic with a {\em standing}
downstream front just at the end of the ramp.
With an observed outflow 
$\tilde{Q}_{\rm out} \le Q_{\rm out}$ from synchronized traffic 
(with $\tilde{Q}_{\rm out} \approx Q_{\rm out}$), 
the average flow upstream is given by
\begin{equation}
\label{qsync}
Q_{\rm sync} = \tilde{Q}_{\rm out} - Q_{\rm rmp}/n \, .
\end{equation}
Hence, the congested region upstream of the on-ramp is growing until
the flow $Q_{\rm main}$ from the main road drops below
$Q_{\rm sync}$. Only then, the congested region starts to ``melt'', 
and it can easily take an additional hour or so until traffic returns to free
flow, even if the maximum flow $Q_{\rm max}$ has never been exceeded
by $(Q_{\rm main} + Q_{\rm rmp}/n)$!

Now, consider the density $\rho_{\rm sync}$ defined by
$Q_{\rm sync} = Q_{\rm e}(\rho_{\rm sync})$ in the congested part of 
the equilibrium flow-density relation $Q_{\rm e}(\rho)$
(i.e. behind its maximum). 
If homogeneous traffic is (me\-\mbox{ta-)}stable 
at $\rho_{\rm sync}$, the on-ramp 
induces ``synchronized'' congested traffic. 
The restriction $Q_{\rm rmp} 
\le \tilde{Q}_{\rm out}/2$ \cite{infl} implies
$Q_{\rm sync} \ge \tilde{Q}_{\rm out}(1 - \frac{1}{2n})$ and
$\rho_{\rm sync} < \rho_{\rm jam}$, so that
``synchronized'' congested flow is significantly higher than the flow
inside traffic jams. 

\subsection{Phase Diagram of Traffic States at Bottlenecks} \label{PD}

\begin{figure}[htbp]
\begin{center}
\includegraphics[width=\textwidth]{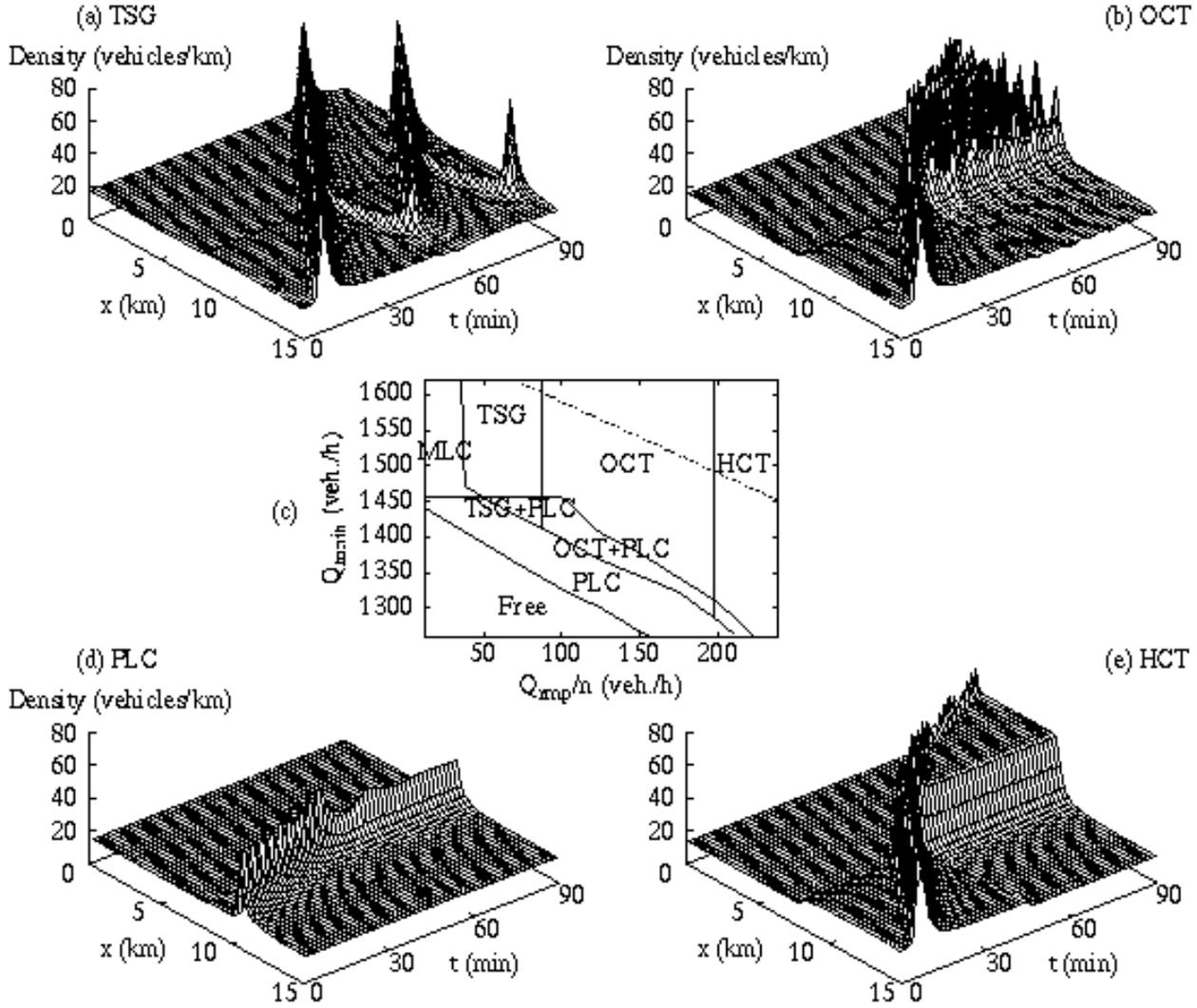}
\end{center}
\vspace*{-5mm}
\caption[]{\label{states}
(a), (b), (d), (e) 
Spatio-temporal dynamics of typical representatives of congested states
which are triggered at an on-ramp by a fully developed perturbation 
travelling upstream. The middle of the on-ramp is located at $x=8.0$\ km. 
The respective states are
(a) triggered stop-and-go traffic (TSG) for
$Q_{\rm main}=1660$~vehicles/h/lane and $Q_{\rm rmp}/n=75$~vehicles/h, 
(b) oscillatory congested traffic (OCT) for
$Q_{\rm main}=1540$~vehicles/h/lane and $Q_{\rm rmp}/n=170$~vehicles/h, 
(d) a pinned localized cluster (PLC) for
$Q_{\rm main}=1450$~vehicles/h/lane and $Q_{\rm rmp}/n=60$~vehicles/h, and
(e) homogeneous congested traffic (HCT) for
$Q_{\rm main}=1350$~vehicles/h/lane and $Q_{\rm rmp}/n=400$~vehicles/h.
The other model parameters used in the above simulations are 
$V_0=110$~km/h, $\rho_{\rm max}=140$~Fz/km, $\tau=40$~s, $T=1.7$~s, $\gamma=1.2$,
$A_0=0.008$, $\Delta A=0.02$, $\rho_c=0.27\rho_{\rm max}$,
$\Delta\rho=0.1\rho_{\rm max}$, and $L=400$~m. They can be considered
as typical for German freeways, where all the above congested traffic
states have been empirically observed.\\
(c) 
Corresponding phase diagram of the traffic states forming in the vicinity of
an on-ramp as a function of the 
inflows $Q_{\rm main}$ and $Q_{\rm rmp}/n$ per freeway lane
on the main road and the 
on-ramp for the above parameters. 
The different states are classified at
$t=90$~min, i.e., after a sufficiently long transient period.
Displayed are homogeneous congested traffic (HCT),
oscillatory congested traffic (OCT),
triggered stop-and-go traffic (TSG),
moving localized clusters (MLC),
pinned localized clusters (PLC),
and free traffic (FT). Spatially coexisting states are indicated
with a plus sign.
The broken line represents the condition
$Q_{\rm main} + Q_{\rm rmp}/n = Q_{\rm max}$ that separates 
the region, in which a breakdown of traffic flow is caused by
exceeding the theoretically possible freeway capacity $Q_{\rm max}$
(upper right corner). Note that
all congested traffic states below this line are caused
by perturbations and, therefore, avoidable by technical control measures.
}
\end{figure}

Close to bottlenecks by on-ramps, lane closures,
gradients, or other effects (including building sites along the road, 
bad street conditions, restricted visibility,
curves, accidents in the opposite lanes, and slow 
vehicles), there can develop a variety of different traffic 
states. Free traffic (FT) and the above discussed
type of ``synchronized'' congested traffic, which we will also call
``homogeneous congested traffic'' (HCT), are 
only two of them. Other congested traffic states that can emerge are
``oscillating congested traffic'' (OCT), ``triggered stop-and-go waves''
(TSG), ``moving localized clusters'' (MLC) and ``pinned (standing) 
localized clusters'' (PLC) (see Fig.~\ref{states}). 
All these different congested states are
observable in empirical traffic data
\cite{kerner-rehb96,kerner-rehb96-2,kerner-sync,emp-Hel,OCT,Daganzo},
although the ``synchronized'' congested states
have sometimes been interpreted differently in the beginning. 
\begin{center}
{\em Insert Figure \ref{states} about here.}
\end{center}
\par
In our traffic simulations, we found out that most of the congested
traffic states occur below the capacity limit $Q_{\rm max}$, and they
are triggered by perturbations in traffic flow. This is only possible since
the outflow $Q_{\rm out}$ from traffic jams is an additional, self-organized
capacity limit, which is smaller than $Q_{\rm max}$ 
(due to the larger time headways
occuring in accelerating traffic). The same fact brings about
that most breakdowns of traffic flow could
be avoided by control measures which manage to 
suppress these perturbations.
\par
Our simulations also indicate that the conditions, under which the different
traffic states appear, are clearly defined \cite{HT-Science,Phase}.
This is illustrated by the phase diagram
in Figure~\ref{states}(c), displaying the traffic states as a function of 
the main flow $Q_{\rm main}$ 
and the on-ramp flow per freeway lane, $Q_{\rm rmp}/n$  
(which is a measure of the flow reduction by a bottleneck).
It explains why usually not all possible congested
traffic states are observed at every bottleneck, but only a typical, 
site-specific subset of them. The phase diagram also allows to understand the
various observed transitions between the traffic states, which
correspond to crossing the phase boundaries separating the different
states. Notice that these phase boundaries can be calculated analytically
in terms of the flows $Q_{{\rm c}i}= Q_{\rm e}(\rho_{{\rm c}i})$
characterizing the stability diagram (see Sec.~\ref{properties}).
This implies that the general structure of the phase diagram
in Fig.~\ref{states}(c) 
is expected to be {\em universal} for all microscopic and macroscopic traffic
models which possess the same instability diagram (i.e. the sequence of
stable, metastable, unstable, convectively stable, metastable, and
stable traffic with increasing density). Only the exact position and the
shape of the phase boundaries will vary from one model to another.

\begin{figure}
\begin{center}
\includegraphics[width=160mm]{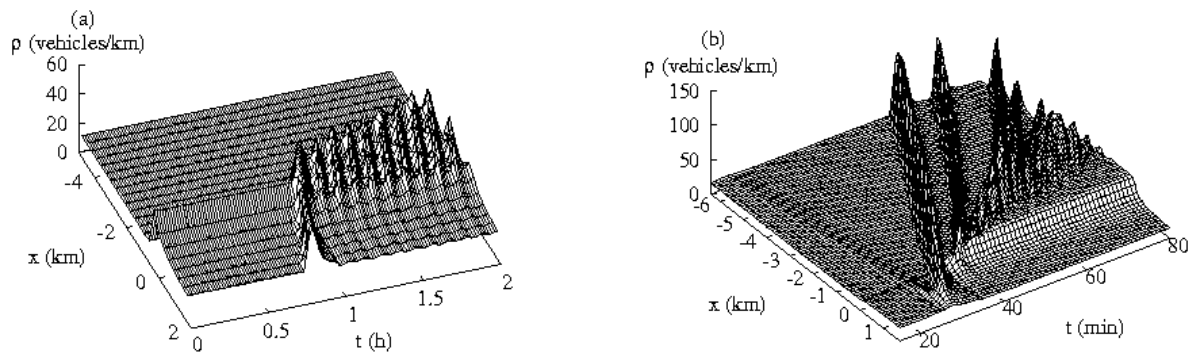}
\end{center}
\caption[]{
\label{tristab}
(a) Metastable pinned localized cluster (PLC) 
and stable oscillating congested traffic (OCT), simulated with the
non-local GKT model (with model parameters 
$v_0=120$ km/h,
$T=1.8$ s, $\tau=50$ s, $\rho_{\rm max}=130$ vehicles/km,
$\gamma=1.2$, $A_0=0.008$, and
$\Delta A=0.008$).
The inflow to the main road is 1400 vehicles per hour
and lane, and the assumed inhomogeneity 
of traffic flow comes from an on-ramp 
of length 500 m with a constant inflow of 140 vehicles per hour
and lane. 
The metastable PLC is triggered by a triangular-shaped density peak 
in the initial conditions (of total width 500~m,
centered at $x=0$), in which the 
density rises from 14 vehicles/km to 40 vehicles/km.
The OCT is triggered by a density wave introduced by 
the downstream boundary conditions. 
\label{PRE}
(b) Spatiotemporal evolution of the traffic density according to the
gas-kinetic-based traffic model \protect\cite{GKT}.
The assumed inhomogeneity of traffic flow comes from an on-ramp 
of length 200 m with an inflow of 220 vehicles per hour
and freeway lane. The inflow to the main road is 
1570 vehicles per hour and lane, and
the breakdown of traffic flow is triggered
by a perturbation $\Delta Q(t)$ of the inflow with a flow peak of 125
vehicles per hour and lane (see main text). 
The assumed model parameters are $V_0=$ 120 km/h, 
$T=1.5$ s,
$\tau=30$ s,
$\rho_{\rm max}=$ 120 vehicles/km, and
$\gamma=1.2$, while
the parameters for the variance prefactor \protect\cite{GKT} 
$A(\rho) = A_0 + \Delta A \{\tanh [ (\rho-\rho_{\rm c})/\Delta \rho] + 1\}$
are $A_0=0.008$,
$\Delta A = 0.02$,
$\rho_{\rm c} = 0.27 \rho_{\rm max}$, and
$\Delta \rho = 0.1 \rho_{\rm max}$.
In order to have a large region of linearly unstable but convectively
stable traffic, we introduced
a ``resignation effect'', i.e., a density-dependent reduction of the
desired velocity $V_0$ to
$V'_0(\rho) = V_0 - \Delta V / \{1 + \exp[(\rho'_{\rm c}-\rho)/\Delta \rho']\}$
with
$\Delta V = 0.9 V_0$,
$\rho'_{\rm c} = 0.45 \rho_{\rm max}$, and
$\Delta \rho = 0.1 \rho_{\rm max}$.
}
\end{figure}

\paragraph{Multistability and Spatial Coexistence of States.}
Note that the phase diagram in Fig.~\ref{states}(c) is for 
fully developed perturbations. 
For smaller perturbations, the phase boundaries will
be shifted. The MLC and PLC regimes may even disappear, leading 
to the boundaries FT$\leftrightarrow$TSG, FT$\leftrightarrow$OCT, 
and FT$\leftrightarrow$HCT. In any case, there 
are regions of multistability, where the resulting traffic state depends
on the initial condition. 
Similar to Lee {\em et al.} \cite{Lee99},
we could even find a small region of tristable traffic (TRI), 
where we can have FT, PLC, or OCT, depending on the the respective
``history'' of the traffic dynamics. Figure~\ref{tristab}(a) shows, for
example, a transition from PLC to OCT.
\par
Another interesting point is the possible spatial coexistence of different
traffic states. If congested traffic is linearly unstable, but
convectively stable, upstream of a bottleneck we can have the
following sequences of states, 
depending on the model parameters and the length of the congested
region: (i) Empirical data and simulation results of 
the sequence HCT$\rightarrow$OCT (without mergers of small density
oscillations to fully developed stop-and-go waves) are presented
in Ref.~\cite{opus}.
(ii) The sequence HCT$\rightarrow$OCT$\rightarrow$TSG (see Fig.~\ref{PRE}(b)) 
is discussed in Ref.~\cite{coexist} (for a slightly modified
GKT model with frustration effects), and we think that it can
explain the empirically observed sequence
``synchronized'' traffic$\rightarrow$``pinch 
region''$\rightarrow$stop-and-go waves reported in Ref.~\cite{Kerner-wide}.
In Sec.~\ref{sec:phase}, we will come back to the
subjects of this paragraph.
\begin{center}
{\em Insert Figure \ref{PRE} about here.}
\end{center}

\paragraph{Qualitative Interpretation of the Transitions.}
Finally, we give a qualitative interpretation of the transitions
between the various congested traffic states and the respective
conditions, under which they appear. For example, the 
transition HCT$\rightarrow$OCT from homogeneous to oscillating
congested traffic occurs, when the ramp flow $Q_{\rm rmp}$ is
decreased so that the synchronized flow 
(\ref{qsync}) falls into the unstable density regime instead of
the (meta-)stable one. A further decrease in the ramp flow will reduce
the average density in the congested region so much that we will 
have an alternation between regimes of free and congested traffic,
which defines stop-and-go waves. We call them triggered stop-and-go
waves (TSG) because, whenever an upstream travelling jam passes the
bottleneck, it triggers a small perturbation which travels downstream as
long as it is small, but eventually changes its speed as it grows and
changes its propagation direction. Finally, the developed perturbation 
travels upstream and passes the bottleneck, where it gives birth to a new
perturbation, and so on. 
\par
Obviously, the small triggered perturbation cannot grow,
when the traffic flow downstream of the bottleneck is (meta-)stable, which 
will lead to a single localized cluster (LC). 
The localized cluster will pass 
the bottleneck in upstream direction (MLC),
if traffic flow is unstable or metastable, there. 
In contrast, the localized cluster 
will be pinned at the bottleneck (PLC), if traffic in the upstream region
is stable, since a density cluster could not survive, there. If the
traffic flow downstream of the bottleneck is also stable, only
free traffic (FT) can persist. Finally, the 
pinned localized cluster will grow to an extended region of congested traffic, 
if its outflow $\tilde{Q}_{\rm out}$ is
smaller than the total flow $(Q_{\rm main}+Q_{\rm rmp}/n)$ at the
bottleneck. 
\paragraph{Discussion.}
Although there are also other proposals for an explanation of the various
transitions between free traffic, congested traffic, and stop-and-go waves
\cite{Kerner-wide},
we see the following advantages of our approach: The transitions result
{\em naturally} from a model for traffic flow on {\em homogeneous} roads,
just by adding source terms (in case of on-ramp
flows) \cite{Phase} or other kinds of inhomogeneities like
speed limits (see Sec.~\ref{sec:phase}), which are {\em
known} to exist. The implementation of these inhomogeneities is {\em
straightforward,} without the requirement of additional refined model
ingredients. Moreover, analogous to the other parameters of the model, the
inhomogeneities 
are easily measurable quantities. There is no model
ingredient, which could not be relatively easily be verified or falsified
by empirical studies. Furthermore, {\em empirical results} confirming the
existence of a phase diagram are already available
\cite{Lee-emp,opus}. 
Finally, we think that the simulated traffic states
related to inhomogeneities of the road arise {\em so} naturally, that any
explanation of empirical data must take these states into account.

\subsection{Velocity Correlations of Successive Cars\label{velcor}}

\begin{figure}
\begin{center}
\includegraphics[width=\half]{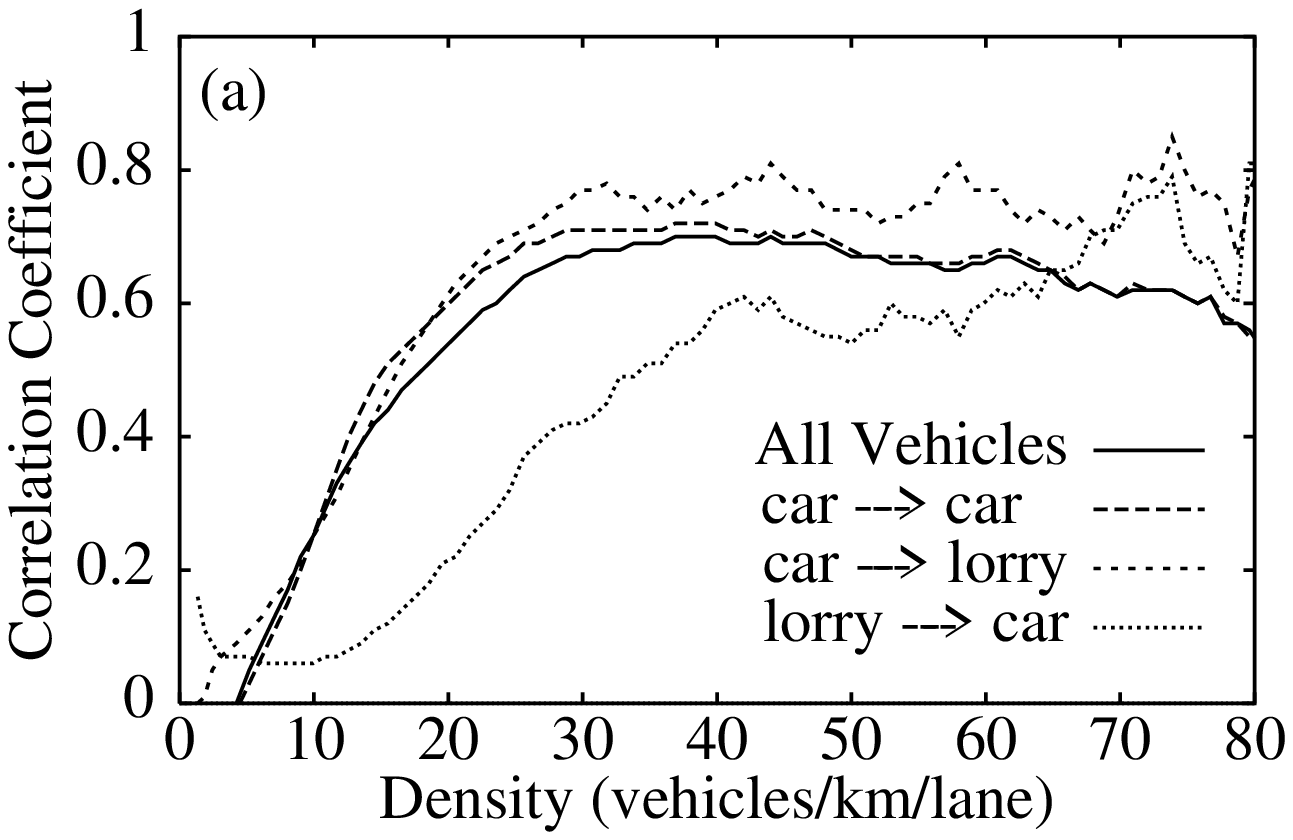}
\includegraphics[width=\half]{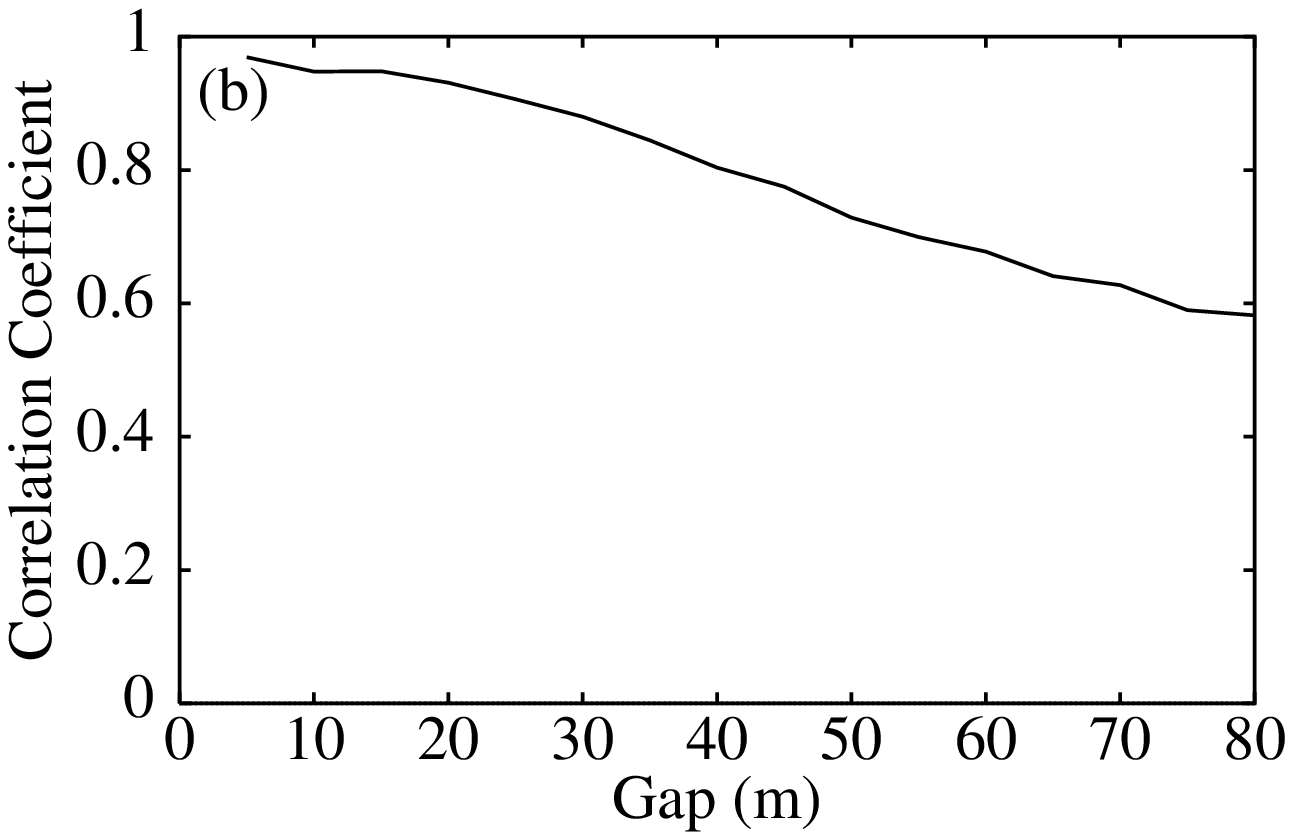}
\end{center} 
\caption[]{\label{COR}Correlation between 
the velocity of successive vehicles 
(a) in the right lane as a function of the density 
(b) as a function of the gap between successive vehicles
for a density of 10 vehicles per kilometer. 
The relations were determined from single-vehicle
data from the Dutch freeway A9. According to our empirical results,
the correlation coefficient is about
zero at small densities and reaches a maximum at around 20
to 30 vehicles per kilometer, where the transition from free to
congested traffic occurs. Afterwards, it stays on a high level
(around 0.65). In the left lane, the velocity correlation is a
little bit higher, probably because of the smaller fraction of lorries.
The correlation coefficient is not sensitive to the type 
of the leading vehicle (car or lorry),
but to that of the following vehicle, which is reasonable. However, even
at small densities, the
correlation depends strongly on the gap to the next vehicle ahead and
becomes almost one for very small gaps. 
This is, because different
velocities would imply a high danger of accidents, then.
}
\end{figure}

With a few exceptions only (see, e.g., Refs.~\cite{nelson,klar-97}),
Boltzmann-like traffic equations are evaluated with the assumption
of ``vehicular chaos'', which neglects correlations $r$ between
the velocities of successive vehicles. This is also true for the
results presented above, mainly, because empirical data of the
correlation coefficient $r$ have not been available until very
recently (see Fig.~\ref{COR}). As expected, the velocity correlations
are small for low vehicle densities, but considerable in congested
traffic. Fortunately, it is possible to generalize the above
non-local, gas-kinetic-based traffic model in a way that allows to
take the velocity correlations into account \cite{mul-HS,transresb}. It turns
out that one has only to replace formula (\ref{SSS}) by
\begin{equation}
 S = \theta  - 2 r \sqrt{\theta \theta'} + \theta' \, . 
\end{equation}
Obviously, this mainly modifies the 
equilibrium velocity a bit, similar to a reduction of velocity 
variances. Therefore, we expect only quantitative, but no
qualitative (fundamental) changes in the simulation
results for finite values of $r$.
\begin{center}
{\em Insert Figure \ref{COR} about here.}
\end{center}

\section{Car-Following Models} \label{CF}

For about fifty years, now, researchers model
freeway traffic by means of continuous-in-time microscopic 
models (car-following models) \cite{Reuschel}.
Since then, a multitude of car-following models have been
proposed, both for single-lane and multi-lane traffic including lane
changes.
\par
In the simplest case, the acceleration of an individual
vehicle depends only on the
distance to the vehicle in front. Well-known models of this type include
the model by Newell \cite{Newell}, or the ``optimal-velocity model'' by
Bando {\it et al.} \cite{Bando}.
To achieve a better anticipative driver behavior and to avoid
collisions, the acceleration in other models 
depends also on the velocity and on the approaching rate
to the front vehicle \cite{Gipps81,Krauss-Diss,Tilch-GFM}.
Besides these simple models intended for basic investigations, 
there are also highly complex ``high-fidelity models'' 
like the Wiedemann model \cite{Wiedemann} or MITSIM \cite{MITSIM}.
They try to reproduce traffic as realistically as possible, but at the 
cost of plenty of parameters, which are difficult to calibrate.
\par
In the following, we will propose and discuss a novel follow-the-leader model
that contains a few parameters only, all of which have a
reasonable interpretation, are known to be relevant, are empirically 
measurable, and have the expected order of magnitude. The
fundamental diagram and the stability properties of the model
can be easily (and separately) calibrated to empirical data, and
simulations of all kinds of observed traffic breakdowns are possible
with the measured boundary conditions \cite{opus}! Moreover,
an equivalent macroscopic version of the model is known, which is
not the case for most other microscopic traffic models. Finally, we
mention that the new model behaves accident-free 
because of the dependence on the relative velocity.
For similar reasons and because of metastability, it shows
the self-organized characteristic traffic constants mentioned before.

\subsection{\label{sec:IDM} The Microscopic Intelligent-Driver Model (IDM)}

The acceleration assumed in the IDM is a continuous function 
of the velocity $v_{\alpha}$, the (netto) gap $s_{\alpha}$,
and the velocity difference (approaching rate)
$\Delta v_{\alpha}$
of vehicle $\alpha$
to the leading vehicle: 
\begin{equation}
\label{IDMv}
\dot{v}_{\alpha} = a^{(\alpha)}
         \left[ 1 -\left( \frac{v_{\alpha}}{v_0^{(\alpha)}} 
                  \right)^{\delta} 
                  -\left( \frac{s_{\alpha}^*(v_{\alpha},\Delta v_{\alpha})}
                                {s_{\alpha}} \right)^2
         \right].
\end{equation}
This expression is a superposition of the acceleration
$a^{(\alpha)}[1-(v_{\alpha}/v_0^{(\alpha)})^{\delta}]$ 
on a free road,
and a braking deceleration
$-a^{(\alpha)}[s_{\alpha}^*(v_{\alpha},\Delta v_{\alpha})/s_{\alpha}]^2$,
describing the interactions
with other vehicles. The deceleration term
depends on the ratio between the ``desired gap'' $s_{\alpha}^*$ 
and the actual gap $s_\alpha$, where the desired gap
\begin{equation}
\label{sstar}
s^*_{\alpha}(v, \Delta v) 
    = s_0^{(\alpha)} + s_1^{(\alpha)} \sqrt{\frac{v}{v_0^{(\alpha)}}}
    + T_{\alpha} v
    + \frac{v \Delta v }  {2\sqrt{a^{(\alpha)} b^{(\alpha)}}}\, ,
\end{equation}
is dynamically varying with the velocity and the approaching rate, reflecting
an intelligent driver behavior.
The IDM parameters are the 
desired velocity $v_0$,
safe time headway $T$,
maximum acceleration $a$,
comfortable deceleration $b$,
acceleration exponent $\delta$,
and the jam distances $s_0$ and $s_1$.
Furthermore, the vehicles have a finite length $l$ 
which, however, has no
dynamical influence.
In Sections \ref{sec:IDM} to \ref{sec:phase}, we will assume
identical ``cars'', while in Sec.~\ref{sec:multi} we assume 
two different types, ``cars'' and ``lorries'', see
Table~\ref{tab:param}.
For better readability, we will drop the vehicle index $\alpha$ in the 
following discussion of the model.
\begin{center}
{\em Insert Table \ref{tab:param} about here.}
\end{center}

\subsection{Equilibrium Traffic of Identical Vehicles}

In equilibrium traffic
($\dot{v}_{\alpha}=0$, $\Delta v_{\alpha}=0$), drivers tend to keep 
a velocity-dependent
equilibrium gap $s_{\rm e}(v_{\alpha})$ to the front vehicle
given by
\begin{equation}
\label{seq}
s_{\rm e}(v) = s^*(v,0) \left[ 1 -
        \left(\frac{v}{v_0}\right)^{\delta}\right]^{-\frac{1}{2}}. 
\end{equation}
In particular, the equilibrium gap of homogeneous 
{\it congested} traffic
($v \ll v_0$) is essentially equal
to the desired gap,
$s_{\rm e}(v) \approx s^*(v,0) = s_0+s_1\sqrt{v/v_0} + v T$, 
i.e., it is
composed of a (small) high-density contribution, and a contribution
$v T$ corresponding to a  time headway $T$.

\begin{figure}[ht]
  \begin{center}
    \includegraphics[width=\textwidth]{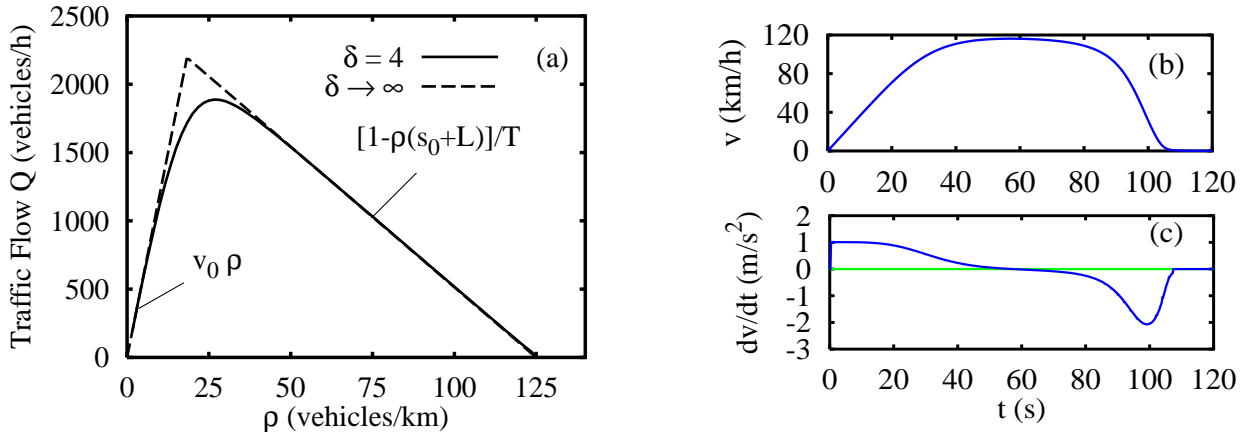}
    \vspace{-5mm}
  \end{center}
  \caption[]{\protect (a) Equilibrium flow-density diagram of identical IDM 
vehicles with variable
acceleration exponent $\delta$ and
the ``car'' parameters of
Table \ref{tab:param} otherwise.
(b), (c) Temporal evolution of the velocity 
and acceleration of a single vehicle
approaching a standing obstacle, which is reached after $t=106$ s
(cf. main text). 
The IDM parameters are $a=1$ m/s$^2$, 
$b=2$ m/s$^2$, and the ``car'' parameters
of Table \ref{tab:param} otherwise.\label{fig:1}}
\end{figure}

Solving Eq. (\ref{seq}) for the equilibrium velocity $v=v_{\rm e}$
leads to simple expressions
only for $s_1=0$ and $\delta=1$, $\delta=2$, or $\delta\to\infty$.
In particular, the equilibium velocity for the special case
$\delta=1$ and $s_0=s_1=0$ is
\begin{equation}
\label{veGKT}
v_{\rm e}(s) 
= \frac{s^2}{2 v_0 T^2} \left(- 1 + \sqrt{1+\frac{4 T^2 v_0^2}{s^2}}
                       \right).
\end{equation}
Macroscopically, homogeneous traffic consisting of identical vehicles can
be characterized by the 
equilibrium traffic flow $Q_{\rm e}(\rho)=\rho V_{\rm e}(\rho)$ 
as a function of the traffic density $\rho$. 
For $\delta=1$ and $s_0=s_1=0$, this ``fundamental diagram'' 
follows from Eq. (\ref{veGKT}) together with the
micro-macro relation between gap and density:
\begin{equation}
\label{srho}
s = \frac{1}{\rho} - l  = \frac{1}{\rho} - \frac{1}{\rho_{\rm max}} \, 
.
\end{equation} 
The resulting relation for the density-dependent equilibrium velocity
is identical with that of the GKT model, 
if the GKT parameter $\Delta A$ is set to zero 
[see Eqs. (\ref{Ve}) and (\ref{tilV})], which is a necessary 
condition for a micro-macro correspondence.
Figure~\ref{fig:1}(a) shows that the acceleration coefficient $\delta$ 
influences the transition region between the free and
congested regimes. For $\delta\to\infty$ and $s_1=0$, 
the fundamental diagram becomes triangular-shaped:
$Q_{\rm e}(\rho) = \mbox{min}(v_0\rho, [1-\rho(l+s_0)]/T)$. 
For decreasing $\delta$, it becomes smoother and smoother.
\begin{center}
{\em Insert Figure \ref{fig:1} about here.}
\end{center}

\subsection{Dynamic Single-Vehicle Properties}

Figures \ref{fig:1}(b) and \ref{fig:1}(c) show how the parameters
$a$ and $b$
determine the acceleration and braking behavior of single vehicles.
At $t=0$, one vehicle starts with zero velocity and 2.5 km of free road
ahead. Initially, it accelerates with $a$ and approaches
smoothly the
desired velocity $v_0$. 
At $x=2.5$ km, we assume a standing obstacle, for example, the end of a
traffic jam. When approaching the obstacle,
the model mimics ``intelligent'' drivers who anticipate
necessary braking decelerations and brake so as not to exceed
the comfortable deceleration $b$ in normal situations \cite{coexist}.

\subsection{Calibration}

The {\it fundamental relations}
of homogeneous traffic are calibrated
with $v_0$ (low density), $\delta$ (transition region),
$T$ (high density), and $s_0$ and $s_1$ (jammed traffic).
The {\it stability behavior} of traffic in the IDM model
is determined mainly by
the model parameters $a$, $b$, and $T$. The density in 
and the outflow from traffic jams are also influenced by $s_0$ and
$s_1$.
Since the accelerations $a$ and $b$ 
do not influence the fundamental diagram,
the model can be calibrated essentially independently with respect to
traffic flows and stability.
As in the GKT model, traffic becomes more unstable for
decreasing $a$ (which corresponds to an increased acceleration time
$\tau=v_0/a$), and for decreasing $T$
(corresponding to reduced safe time headways). 
Furthermore, the instability increases with 
growing $b$. This is also plausible, because an increased 
desired deceleration $b$
corresponds to a less anticipative braking
behavior.

\subsection{Collective Behavior and Stability Diagram}

\begin{figure}[t]
  \begin{center}
    \includegraphics[width=\textwidth]{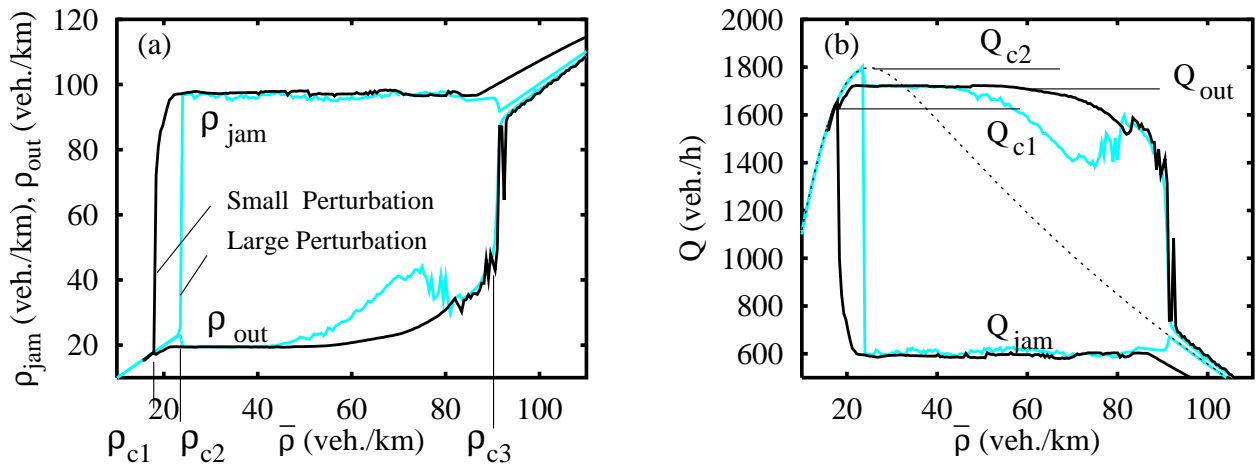}
    \vspace*{-8mm}
  \end{center}
  \caption[]{\protect Stability diagram of homogeneous traffic 
(for ``car'' parameters) in a closed system
as a function of the homogeneous density $\overline{\rho}$
for small (grey) and large (black) initial
perturbations of the density. In plot (a),
the upper two lines display the density inside
of density clusters after a stationary state has been reached. 
The lower two lines represent the density between the clusters.
Plot (b) shows the corresponding flows and the
equilibrium flow-density relation (dotted).
The critical densities $\rho_{{\rm c}i}$ and flows
$Q_{{\rm c}i}$ are discussed in the 
main text.\label{fig:stab}}
\end{figure}

Although we are interested in realistic 
{\it open} systems, it turned out that many features can be explained
in terms of the stability behavior in a {\it closed} system.
Figure \ref{fig:stab}(a) shows the stability diagram of 
homogeneous traffic
on a circular road. The control parameter is the
homogeneous (average) density $\overline{\rho}$. We applied both a very small 
and a large localized perturbation to check for linear and nonlinear
stability, and plotted the resulting minimum
($\rho_{\rm out}$) and maximum ($\rho_{\rm jam}$)
densities after a stationary situation was reached. 
The resulting diagram is very similar to that of the
macroscopic KKKL and GKT models \cite{kk-94,GKT}. 
In particular, it displays the following realistic features:
(i) Traffic is stable for very low and high densities, but unstable for
intermediate densities. 
(ii) There is a density range $\rho_{\rm c1}\le\overline{\rho}\le\rho_{\rm c2}$
of metastability, i.e., only perturbations of sufficiently large
amplitudes grow, while smaller perturbations disappear. Note that, 
for most IDM parameter sets, there is no second metastable range at 
higher densities, in contrast to the GKT and KKKL models. 
(iii) The density inside of traffic jams 
and the associated flow $Q_{\rm jam}=Q_{\rm e}(\rho_{\rm jam})$,
cf. Fig. \ref{fig:stab}(b),
do not depend on
$\overline{\rho}$. As further ``traffic constants'', at least
in the density range 20 veh./km $\le \overline{\rho} \le$ 40 veh./km, 
we observe a constant outflow $Q_{\rm out}=Q_{\rm e}(\rho_{\rm out})$
and propagation velocity 
$v_{\rm g}=(Q_{\rm out} - Q_{\rm jam}) / (\rho_{\rm out}-\rho_{\rm jam})
\approx -15$ km/h of jams.
Figure~\ref{fig:stab}(b) shows the stability diagram for the flows.
In particular, we have $Q_{\rm c1}<Q_{\rm out}<Q_{\rm c2}$,
where $Q_{{\rm c}i}=Q_{\rm e}(\rho_{{\rm c}i})$, i.e., the outflow from
congested traffic is metastable, while in the GKT,
$Q_{\rm out}\approx Q_{\rm c2}$ is only marginally stable.
\begin{center}
{\em Insert Figure \ref{fig:stab} about here.}
\end{center}
\par
In {\it open} systems, a third type of stability becomes relevant.
Traffic is {\it convectively} stable, if, after a sufficiently long time,
all perturbations are convected out of the system.
Both in the macroscopic models and in the IDM, there is a considerable
density region $\rho_{\rm cv}\le \overline{\rho}\le \rho_{\rm c3}$,
where traffic is linearly unstable but convectively stable.

\subsection{\label{sec:mic}Microscopic Implementation of
Bottlenecks}

\begin{figure}[h]
\begin{center}
\includegraphics[width=\textwidth]{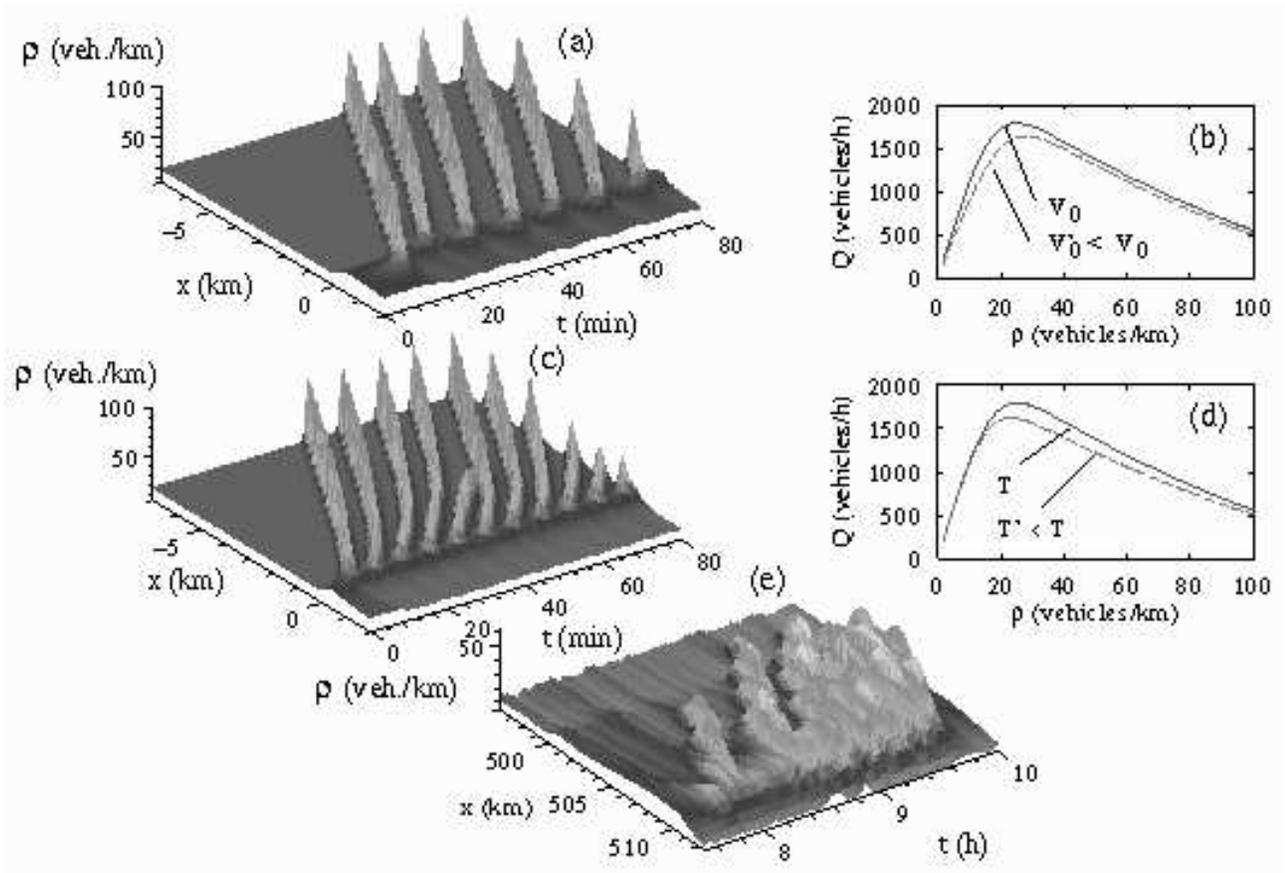}
\end{center}
\vspace*{-5mm}
\caption[]{\protect
Traffic breakdowns at differently implemented bottlenecks.
(a) IDM simulation of the spatio-temporal density 
(inflow $Q_{\rm main}=1670$ vehicles/h)
for a bottleneck corresponding to a
decrease of $v_0$ in the downstream region. The density is derived
from the microscopic distance $s$ via relation (\ref{srho}) and a
subsequent linear interpolation.
(b) Related equilibrium flows
upstream and downstream. 
(c), (d) Bottleneck corresponding to 
an increase of the safe time headway $T$.
(e) Density obtained from 1-min detector data
of the German freeway A9-South on Oct. 29, 1998. The traffic
breakdown takes place upstream of the intersection 
``Neufahrn'' at $x=512$ km.\label{fig:bottleneck}}
\end{figure}

In {\it macroscopic} simulations, a natural implementation of road
inhomogeneities is given by on- and off- ramps, which appear
as a source term in the continuity equation 
for the density.
An explicit {\it microscopic} modelling of ramps, however, would require
a multi-lane model with explicit simulation of lane changes. 
In order to avoid the associated complications, 
one can either apply the micro-macro link
and simulate the ramp section macroscopically (see Sec.~\ref{micro-macro}),
or introduce {\it flow-conserving} inhomogeneities by making one or more
model parameters dependent on the location $x$ of the road.
Suitable parameters for the IDM
are $v_0$ or $T$ \cite{coexist,opus}.
Local parameter variations act as a bottleneck,
if the outflow $Q'_{\rm out}$ from congested traffic in the downstream
section is reduced with
respect to the outflow $Q_{\rm out}$ in the upstream section.
This requires a reduced desired velocity $v'_0<v_0$
or increased time headway $T'>T$, or both.
Figure~\ref{fig:bottleneck}(a) shows a traffic breakdown induced by a linear
decrease of the desired velocity from $v_0=120$ km/h for $x\le -L/2$ to
$v_0'=95$ km/h for $x\ge L/2$,
while Fig.~\ref{fig:bottleneck}(c)
shows the same effect for an increase of the safe time headway
from $T=1.2$ s to $T'=1.45$ s ($L=200$ m).
Both flow-conserving bottlenecks
result in a similar traffic dynamics which, however, depends strongly
on the amplitude of the parameter variation. 
Qualitatively the same dynamics is observed in real traffic data 
[Fig. \ref{fig:bottleneck}(e)] and in macroscopic models including
on-ramps with a ramp flow of
$Q_{\rm rmp}/n=(Q_{\rm out}-Q'_{\rm out})$.
This suggests to define a general ``bottleneck strength''
$\Delta Q$ by
\begin{equation}
\label{deltaQ}
\Delta Q := Q_{\rm rmp}/n + Q_{\rm out}-Q'_{\rm out}. 
\end{equation}
In particular, we have $\Delta Q=Q_{\rm rmp}/n$ for on-ramp bottlenecks,
and $\Delta Q=Q_{\rm out}-Q'_{\rm out}$ for flow-conserving bottlenecks.
In the following, we will vary $v_0$.
The regions with locally decreased desired velocity can be interpreted
as sections with uphill gradients (which reduce the maximum velocities of 
vehicles). 
\begin{center}
{\em Insert Figure \ref{fig:bottleneck} about here.}
\end{center}

\subsection{\label{sec:phase}Phase Diagram of Congested Traffic}

\begin{figure}[h]
  \begin{center}
    \includegraphics[width=\textwidth]{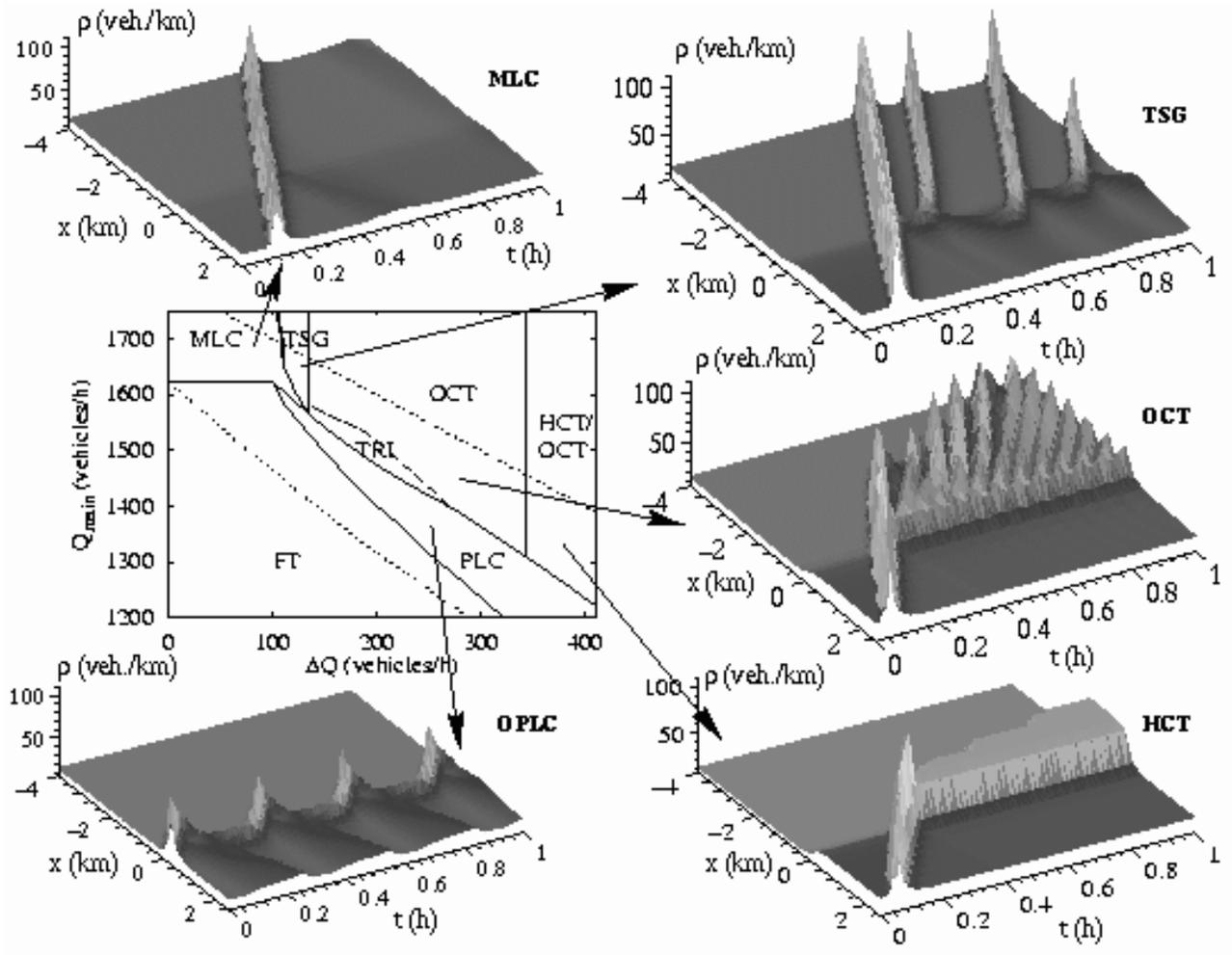}
  \end{center}
\vspace*{-5mm}
\caption[]{\protect 
Phase diagram of congested traffic states and corresponding
spatio-temporal density plots as
simulated with the IDM.
The control parameters are the inflow
$Q_{\rm main}$ and the bottleneck strength $\Delta Q$,
which increases with
$(v_0-v'_0)$. The traffic states FT, HCT, OCT, TSG, MLC, and (O)PLC
are explained in the main text. 
``TRI'' indicates a tristable region.\label{fig:phase}}
\end{figure}

\begin{figure}[h]
  \begin{center}
     \includegraphics[width=0.9\textwidth]{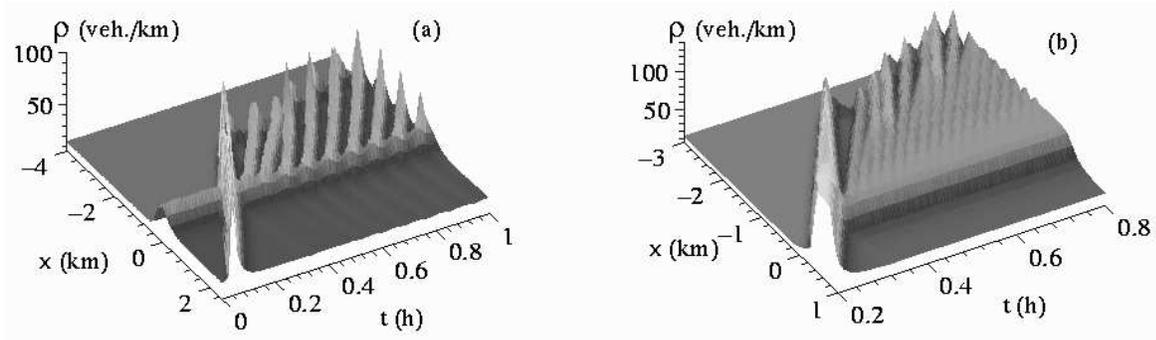}
  \end{center}
  \vspace*{-5mm}
\caption[]{\protect
(a) Transition from a pinned localized cluster (PLC) to oscillating
congested traffic (OCT) in the tristable region
($Q_{\rm main}=1480$ vehicles/h, $v'_0=24$ m/s corresponding to 
$\Delta Q=220$ vehicles/h), triggered by a large perturbation.
(b) Spatial coexistence of HCT and OCT for $Q_{\rm main}=1350$ vehicles/h
and $v'_0=16$ m/s corresponding to $\Delta Q=400$ 
vehicles/h.\label{fig:coextri}}
\end{figure}

In contrast to {\it closed} systems, in which the long-term behavior and
stability is essentially
determined by the average traffic density, the dynamics of 
{\it open} systems
is controlled by the inflow $Q_{\rm main}$.
Furthermore, traffic congestions depend on road inhomogeneities and,
because of hysteresis effects, on the history of
previous perturbations.
For a given history, the traffic states can be summarized by a phase
diagram spanned by $Q_{\rm main}$ and
$\Delta Q$.
Figure~\ref{fig:phase} shows the IDM phase diagram for traffic
states that develop after a single density cluster
crosses the inhomogeneity. 
Depending on $Q_{\rm main}$ and $\Delta Q$, the initial perturbation
(i) dissipates, resulting in free traffic (FT), (ii) travels through the
inhomogeneity as a moving localized cluster (MLC)
and neither dissipates nor triggers new breakdowns,
(iii) triggers a traffic
breakdown to a pinned localized cluster (PLC), which remains 
localized near the inhomogeneity for 
all times and either is stationary,
cf. Fig.~\protect\ref{fig:coextri}(a) for $t<0.2$ h,
or  oscillatory (OPLC).
(iv) Finally, the initial perturbation can induce 
extended congested traffic (CT), whose downstream boundaries
are fixed at the inhomogeneity, while the upstream front propagates
further upstream in the course of time. This kind of congested traffic 
can be homogeneous (HCT), oscillatory (OCT), or consist of
triggered stop-and-go waves (TSG).
The observed traffic states and their
qualitative dependence on $Q_{\rm main}$ and $\Delta Q$
is the same as for the GKT model, Fig. \ref{states} when identifying
$\Delta Q$ with $Q_{\rm rmp}/n$ according to Eq. (\ref{deltaQ}).
In contrast to OCT, where there is
permanently congested traffic at the inhomogeneity (``pinch region''
\cite{Kerner-wide,coexist}), the TSG state is characterized by a
series of isolated density clusters, each of which triggers a new
cluster as it passes the inhomogeneity. 
\begin{center}
{\em Insert Figure \ref{fig:phase} about here.}
\end{center}

\paragraph{Boundaries between and Coexistence of Traffic States.}
Simulations show that the outflow $\tilde{Q}_{\rm out}$
from the nearly stationary downstream fronts of OCT and HCT
satisfies $\tilde{Q}_{\rm out}\le Q'_{\rm out}$, where
$Q'_{\rm out}$ is the outflow from clusters in homogeneous systems for
the downstream model parameters. If the bottleneck is not too strong,
we have $\tilde{Q}_{\rm out}\approx Q'_{\rm out}$.
Then, for all types of bottlenecks, the congested traffic flow is
given by
$Q_{\rm cong}=\tilde{Q}_{\rm out} - Q_{\rm rmp}/n \approx 
Q'_{\rm out}  - Q_{\rm rmp}/n$, or
\begin{equation}
\label{Qcong}
Q_{\rm cong} 
\approx Q_{\rm out} - \Delta Q \, ,
\end{equation}
which generalizes (\ref{qsync}).
Extended congested traffic (CT) 
only persists, if the inflow exceeds the congested
traffic flow. Otherwise, it dissolves to PLC.
This gives the boundary
\begin{equation}
\label{CT-PLC}
\mbox{CT}\to\mbox{PLC}: \Delta Q\approx Q_{\rm out}-Q_{\rm main}.
\end{equation}
If congested traffic flow is {\it convectively} unstable,
the resulting oscillations lead to TSG or OCT.
If it is {\it linearly} stable, 
$Q_{\rm cong} < Q_{\rm c3}$, we have HCT.
If it is convectively stable, but linearly unstable,
$Q_{\rm cong} \in [Q_{\rm c3},Q_{\rm cv}]$, one has a spatial
{\it coexistence} of states with HCT near the bottleneck and
OCT further upstream \cite{coexist} (Fig.~\ref{fig:coextri}), which is
frequently found in empirical data of congested traffic.
In the IDM, this frequent occurrence is reflected by the 
wide range of flows falling into this regime.
For the ``car'' parameters, we have
$Q_{\rm c3}=600$ vehicles/h and $Q_{\rm cv}=1340$ vehicles/h.
\begin{center}
{\em Insert Figure \ref{fig:coextri} about here.}
\end{center}

\paragraph{Pinch Effect and Merging of Clusters.}

Careful investigations of traffic data related to OCT states 
\cite{Kerner-wide} showed two phenomena:
(i) In a narrow region near the inhomogeneity, congested traffic is
nearly stationary (``pinch region''),
while further upstream, there are oscillations of the traffic
density.
(ii) While propagating upstream, the oscillations grow and merge to
a few large-amplitude density clusters with free traffic in between.
Detector data of other freeways,
however, show a pinch effect without mergers, 
cf. Fig. \ref{fig:bottleneck}(e). 
Simulating the IDM with the ``car'' parameters leads to very few
mergers, cf. Figs. \ref{fig:bottleneck}(c) and \ref{fig:phase}.
For other parameters, however, the IDM reproduces mergers ending up with
stop-and-go traffic \cite{coexist}.
A possible explanation is the {\it ``starvation effect''}:
For the parameters chosen in this article,
the outflow $Q_{\rm out}$ from density
clusters is in the middle of the metastable region
[Fig. \ref{fig:stab}(b)], so medium-sized and large density clusters
persist. For the parameters of Ref. \cite{coexist}, however,
the outflow from density clusters satisfies
$Q_{\rm out}\approx Q_{\rm c1}$, so only  
large-amplitude clusters survive, while all others dissipate.

\paragraph{Multistability.}

In general, the local phase
transitions between free traffic,
pinned localized states, and extended congested states are hysteretic. 
In the regions between the two dotted lines of the
phase diagram in Fig.~\ref{fig:phase}, both, free and congested traffic is 
possible, depending on the previous history. 
In particular, 
for all five indicated phase points (but not for the simulations of
Fig. \ref{fig:bottleneck}),
free traffic would persist without the downstream perturbation. In contrast,
the transitions PLC$\longrightarrow$OPLC, and
HCT$\longrightarrow$OCT$\longrightarrow$TSG 
seem to be non-hysteretic, i.e., the type of pinned
localized cluster or of extended congested traffic, is uniquely
determined by $Q_{\rm main}$ and $\Delta Q$.

In a small subset of the metastable region labelled ``TRI'' in
Fig. \ref{fig:phase}, we even found {\it tristability} between FT,
PLC, and OCT. We obtained qualitatively the same also for the GKT model,
and it has been found for the KKKL model
with OPLC instead of PLC for the pinned localized state  \cite{Lee}.
Figure \ref{fig:coextri}(a) shows that
a single moving localized cluster passing the inhomogeneity
triggers a transition from PLC to OCT. Starting with free traffic,
the same perturbation would trigger OCT as well, while we never found reverse
transitions OCT $\to$ PLC or OCT $\to$ FT (without a reduction of the 
inflow). That is, FT and PLC are metastable in the tristable region, 
while OCT is stable.

\subsection{Multi-Species Single-Lane Traffic}\label{sec:multi}

\begin{figure}[h]
  \begin{center}
    \includegraphics[width=1\textwidth]{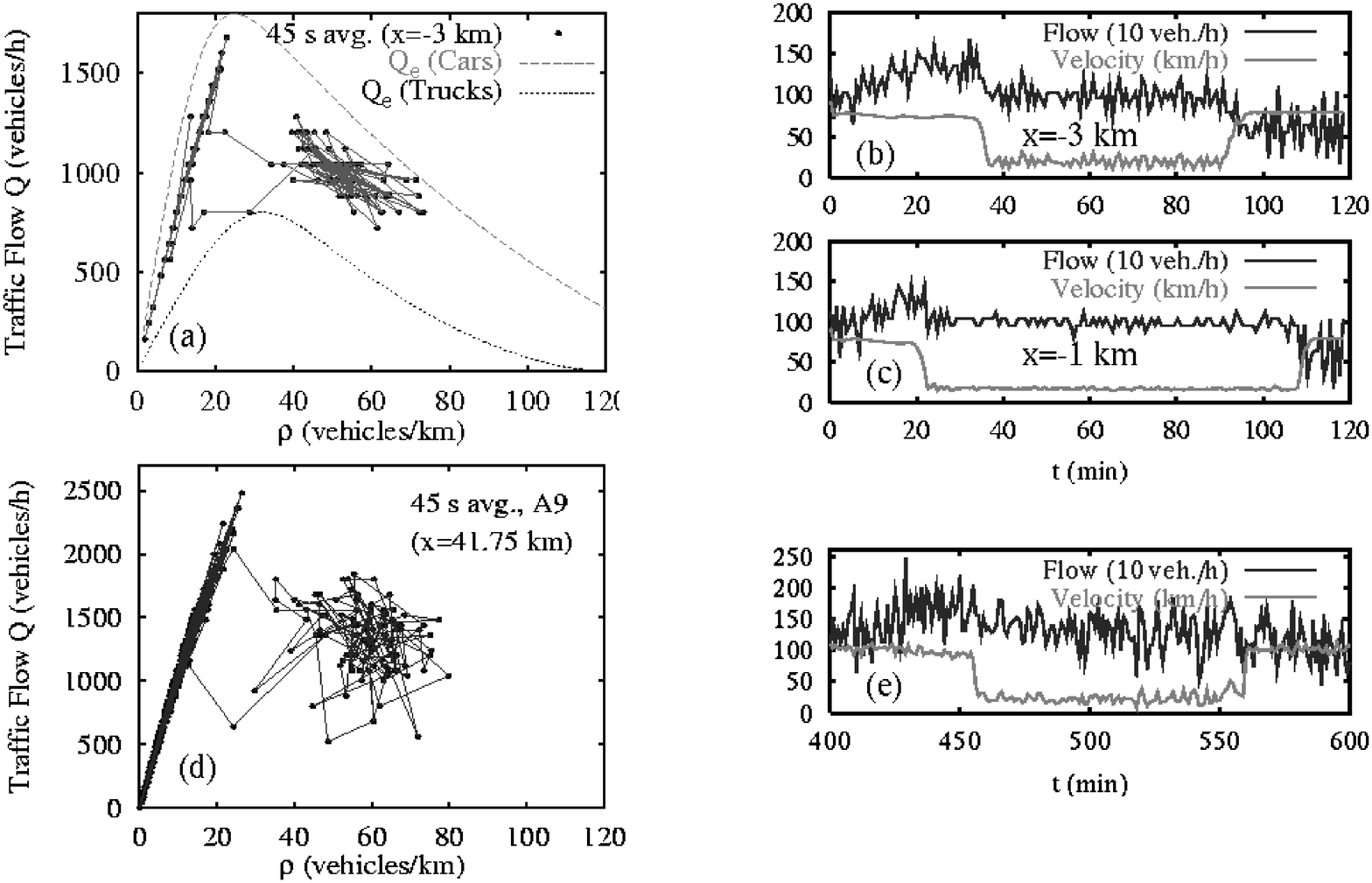}
  \end{center}
\vspace*{-5mm}
\caption[]{\protect
(a) Flow-density diagram and (b),(c) time series of single-lane
heterogeneous traffic.
(d),(e) Empirical data of extended
congested traffic on the Dutch freeway A9.\label{fig:scatter}}
\end{figure}

In this section, we assume heterogeneous single-lane
traffic consisting of 70\%
``cars'' and 30\% ``lorries'',
where the latter are
characterized by a lower desired velocity, lower accelerations, and a larger
safe time headway compared to cars (Table~\ref{tab:param}).
Again, we simulate a traffic breakdown to HCT at a flow-conserving
inhomogeneity, where the desired velocity of cars is
reduced from 120 km/h to 65 km/h, and that of lorries from 
80 km/h to 36 km/h. (Similar results are found for less dramatic
increases of $T$.)
To compare the result with real traffic data, we implement
``virtual'' detectors at several fixed locations.
The detectors record passage times and velocities
of each vehicle to determine the macroscopic flow
$Q=n_{\tau}/\tau$
(where $n_{\tau}$ is the number of passing vehicles
in the averaging interval $\tau=45$ s), 
the arithmetic velocity average $V$, 
and the density $\rho=Q/V$.
Figure~\ref{fig:scatter} shows the resulting fundamental diagram
3 km upstream of the bottleneck,
and time series of $Q$ and $V$ at three
upstream locations.
For comparison,
Fig.~\ref{fig:scatter}(d) and (e) show
45 s averages of {\it real}
single-vehicle data of the Dutch freeway A9
from Haarlem to Amsterdam on October 14, 1994. The detector
is about 0.7 km
upstream of the on-ramp causing the traffic breakdown.
The simulated and real
traffic data agree qualitatively, in particular in the following
respects:
(i) There is a wide scattering of flow-density data in the congested
regime (looking like an anisotropic two-dimensional random walk), 
while the data occupy a nearly one-dimensional region in the
free regime. 
(ii) 
The distribution of flow-density data 
shows the typical inverse-$\lambda$ form with
a distinct gap between free and congested traffic data.
(iii) 
During the breakdown, the velocity drops to 10-20 km/h, while
the flow is reduced by only about 20\%.
(iv) 
In all regions, the relative fluctuations of the velocity are
smaller than those of the flow. Near the bottleneck, the fluctuations
of congested traffic flow are much smaller than those of free traffic,
while further upstream, the fluctuations grow. 
\begin{center}
{\em Insert Figure \ref{fig:scatter} about here.}
\end{center}
\par
Notice that the fluctuations of the simulated data
come essentially from the different vehicle types
and not from deterministic instabilities. Macroscopically, the
situation of Fig.~\ref{fig:scatter}
corresponds to a spatial coexistence of HCT and OCT. 
(For a pure OCT or TSG state, there would be no gap
between the flow-density data of free and congested traffic, both in
real traffic data and in simulations \cite{opus}.)

\section{Micro-Macro Link} \label{micro-macro}

One of the biggest problems in
comparing microscopic and macroscopic models via coarse graining
is that a separation between the microscopic scale (of single vehicles) and
the macroscopic scale (allowing to determine densities and average
velocities) is hardly possible.
Either the averaging intervals are too small and
aggregate quantities like the density cannot be defined consistently,
or the averaging intervals are too large so that the
dynamics of the model is smoothed out.
\par
Generalizing an idea sketched in Ref.~\cite{parisi}, we will now
propose a way of obtaining macroscopic from microscopic traffic
models, which is different from the gas-kinetic approach and (at
least) applicable to the case of identical driver-vehicle units. For this, we
define an average velocity by linear interpolation between 
the velocities of the single vehicles $\alpha$:
\begin{equation}
V(x,t)=\frac{v_{\alpha}(t)[x_{\alpha-1}(t)-x]
+v_{\alpha-1}(t)[x-x_{\alpha}(t)]}
{x_{\alpha-1}(t)-x_{\alpha}(t)} \, ,
\end{equation}
where $x_{\alpha-1}\ge x \ge x_{\alpha}$, and the index $\alpha$ of
driver-vehicle units increases against the driving direction.
While the derivative with respect to $x$ gives us 
\begin{equation}
 \frac{\partial V}{\partial x} 
 = [v_{\alpha-1}(t)-v_{\alpha}(t)]/[x_{\alpha-1}(t)-x_{\alpha}(t)] \, ,
\end{equation}
the derivative with respect to $t$ gives us the {\em exact} equation
\begin{equation}
  \left( \frac{\partial}{\partial t} 
 +V\frac{\partial}{\partial x}\right)V =
A(x,t) \, ,
\end{equation}
where
\begin{equation}
A(x,t)=
\frac{a_{\alpha}(t)[x_{\alpha-1}(t)-x]+a_{\alpha-1}(t)[x-x_{\alpha}(t)]}
{x_{\alpha-1}(t)-x_{\alpha}(t)}
\end{equation}
is the linear interpolation of the  
single-vehicle accelerations $a_{\alpha}$ characterizing the microscopic
model. For most car-following models, the acceleration function can 
be written in the form
$a_{\alpha}=a_{\rm mic}(v_{\alpha},\Delta v_{\alpha}, s_{\alpha})$, where
$\Delta v_{\alpha}=(v_{\alpha}-v_{\alpha-1})$ 
is the approaching rate, and $s_{\alpha}=(x_{\alpha-1} -
x_\alpha - l_\alpha)$ the (netto) distance to the vehicle in front.
In order to obtain a macroscopic system of partial differential equations for
the average velocity and density, the arguments $v_{\alpha}$, 
$\Delta v_{\alpha}$ and $s_{\alpha}$
of the  single-vehicle acceleration have to be expressed 
in terms of macroscopic fields as well.
Specifically, we make the following approximations:
$A(x,t)\approx a_{\rm mic}(V(x,t),\Delta V(x,t),S(x,t))$, with
$\Delta V(x,t)=[V(x,t)-V_{\rm a}(x,t)]$, 
$S(x,t)=\frac{1}{2}[\rho^{-1}(x,t)+\rho^{-1}_{\rm a}(x,t)] 
 - \rho_{\rm max}^{-1}$, and
the non-locality given by $g_{\rm a}(x,t) = g(x+1/\rho(x,t),t)$
with $g\in \{\rho,V\}$.
In this way, we obtain a non-local macroscopic velocity equation,
which supplements the continuity equation (\ref{contin}) for the vehicle density
and defines, for a given microscopic traffic model, a complementary 
macroscopic model. Note that it does not contain a pressure term
$(1/\rho)\partial P/\partial x$ with $P=\rho\theta$, 
in contrast to the GKT model described above.
\par
\begin{figure}[t]
\begin{center}
\includegraphics[width=6.4in]{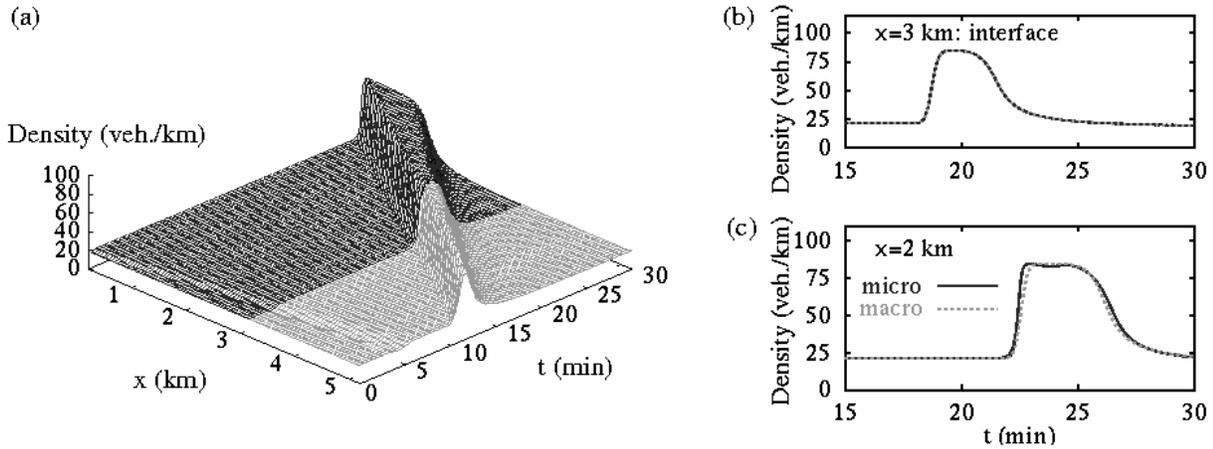}
\end{center}
\caption[]{\protect Results of simulations 
with the microscopic IDM (dark grey) 
and its macroscopic counterpart (light grey), assuming identical
boundary and interface conditions.\label{fig:micmac1}}
\end{figure}

As Fig. \ref{fig:micmac1} shows for the intelligent
driver model, there are 
microscopic models for which the above procedure
leads to a remarkable agreement. 
First, we have simulated an open freeway stretch with the macroscopic
counterpart of the IDM with given macroscopic initial and boundary
conditions. Then, we have simulated the first three kilometers of the
stretch microscopically with the IDM [Fig.~\ref{fig:micmac1}(a), dark grey], 
using the same upstream boundary
condition. As downstream boundary condition, we used the macroscopic
simulation result at the interface ($x=3$\,km) to the
remaining macroscopic simulation section [Fig.~\ref{fig:micmac1}(a),
light grey]. The density profile at the interface is illustrated in
Figure~\ref{fig:micmac1}(b). Figure~\ref{fig:micmac1}(c) 
shows the density profile
1 km upstream of the interface obtained by the microscopic 
simulation, compared with that
obtained from the separate macroscopic simulation of the whole system.
The density $\rho(x,t)$
in the microscopic simulation section 
for $x_{\alpha-1} \ge x \ge x_{\alpha}$ was determined via the 
formula 
\begin{equation}
 \frac{1}{\rho(x,t)}=\frac{1}{\rho_{\rm max}} 
+ \frac
 {s_{\alpha}(t)[x_{\alpha-1}(t)-x]+s_{\alpha-1}(t)[x-x_{\alpha}(t)]}
 {x_{\alpha-1}(t)-x_{\alpha}(t)} \, .
\end{equation}
\begin{center}
{\em Insert Figure \ref{fig:micmac1} about here.}
\end{center}
\par
\begin{figure}
\begin{center}
\includegraphics[width=140mm]{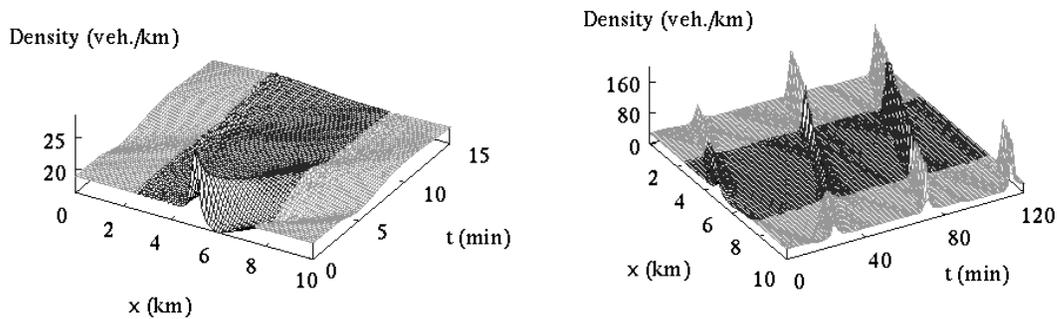}
\end{center}
\caption[]{\protect Results of 
simulations of a circular road, half of which
was simulated microscopically (dark grey), while the other half was
simulated macroscopically (light grey). Whereas the left picture shows the
forward propagation of a decaying initial perturbation at small
vehicle density, the right one illustrates the development of a
backwards moving traffic jam at medium
density.\label{fig:micmac2}}
\end{figure}

The micro-macro link of traffic models is only of practical relevance,
if micro- and macrosimulations can be carried out {\em simultaneously}.
For this, we have derived rules how to translate macroscopic initial 
and boundary conditions into microscopic ones, with almost identical
simulation results. In addition, we have developed methods 
to determine and apply the interface conditions 
(i.e. the respective upstream and downstream
boundary conditions at interfaces) {\em dynamically} during the 
simulation, i.e. {\em on-line}. Since information must be able to
flow through
the interface in both directions, the formulation of dynamic interface
conditions is a particularly tricky task which cannot be discussed, here.
Figures~\ref{fig:micmac2}(a) and 
(b) show examples for 
the spatio-temporal evolution of the density on a circular road.
One half is simulated with the
microscopic IDM (dark grey) and the
other half is simulated with the macroscopic counterpart
of the same model (light grey). As can be seen, it
is possible to connect both sections in a way that 
small perturbations propagating in {\it forward} direction as well as 
developed density clusters propagating {\it backwards} can pass the 
interfaces without any significant changes in the shape or propagation 
velocity.
\begin{center}
{\em Insert Figure \ref{fig:micmac2} about here.}
\end{center}
\par
\begin{figure}
\begin{center}
\includegraphics[width=3.5in]{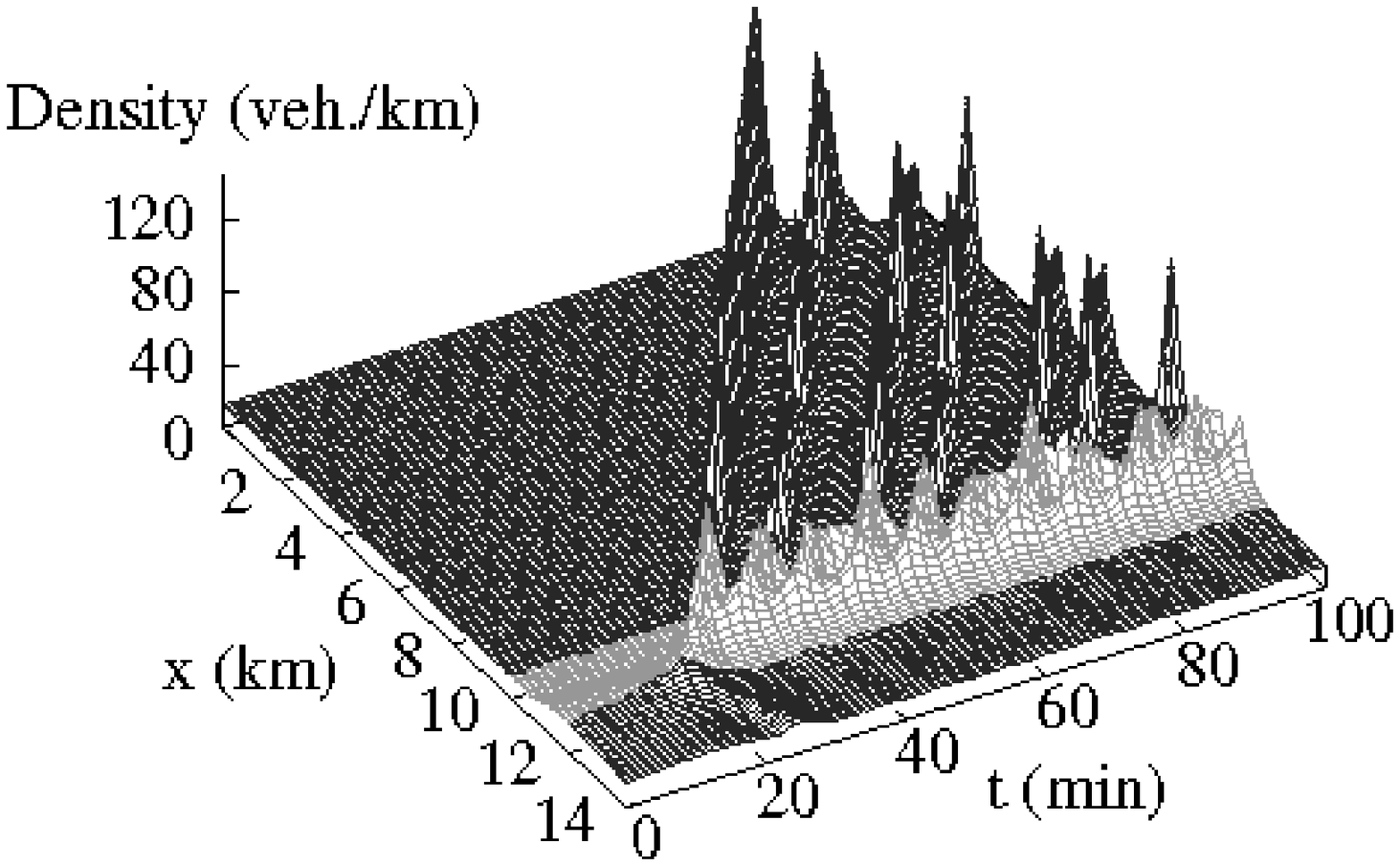}
\caption[]{\protect Result of a simultaneous 
micro-macro-simulation of triggered stop-and-go traffic, which is
caused by high inflow from the ramp at $x=11$\,km. 
While the ramp section (light grey) is implemented macroscopically,
the other two sections (dark grey) were simulated 
microscopically.\label{fig:ramp}}
\end{center}
\end{figure}

While traffic simulations with microscopic single-lane models are more
intuitive and detailled than macroscopic simulations, a microscopic 
implementation of on- and off-ramps would require rather complex 
multi-lane models with lane-changing rules. Ramps can be treated much
easier by macroscopic models, where they just enter 
by a simple source term in the continuity equation. Hence,
the micro-macro link can be applied to combine both advantages.
Figure~\ref{fig:ramp} shows an open system that is simulated 
essentially with a microscopic single-lane model (dark grey).
Only the ramp section (light grey) is simulated macroscopically.
The on-ramp flow triggers a traffic breakdown to a triggered
stop-and-go state, i.e. a cascade
of traffic jams which propagate in the upstream microscopic section.
The qualitative features of this complicated dynamics are in very good
agreement with a purely macroscopic simulation of the same system.
\begin{center}
{\em Insert Figure \ref{fig:ramp} about here.}
\end{center}

\section{Summary and Conclusions}

In this paper, we have shown that all presently known macroscopic
phenomena of freeway traffic, including (i) the fundamental diagram,
(ii) the characteristic parameters of congested traffic
(outflow, propagation velocity, etc.), and 
(iii) the transitions between free traffic,
``synchronized'' or other congested traffic, and stop-and-go traffic
can be reproduced and explained by microscopic and macroscopic
traffic models based on plausible assumptions and realistic
parameters. (iv) These models have also successfully reproduced a variety of
empirical observations in a semi-quantitative way, using the measured
boundary conditions \cite{GKT-scatter,opus}. 
(v) Moreover, simply by simulating a mixture of
different vehicle types, 
we obtained the observed scattering of congested
flow-density data, although our
traffic models are deterministic and have unique equilibrium
relations. (vi) Generalizations to multi-lane
traffic \cite{mul-HS,hoog} are expected to
result in an even better agreement with empirical findings, 
e.g., a larger scattering of
``virtual'' detector data and flow-stabilizing
effects of speed limits \cite{Applet-engl,Sollacher-control}.
(vii) Finally, we established a link between microscopic and
macroscopic models, which allows to carry out simultaneous
micro- and macrosimulations of neighboring freeway sections. 
To us, apart from developing measures for traffic optimization,
the most interesting open question is which kind of observable
phenomena are produced by the heterogeneity of
driver-vehicle units {\em in addition} to the scattering of traffic
data (see Ref.~\cite{helb-nat} for an example).

\subsection*{Acknowledgments}

The authors want to thank for financial support by the BMBF (research
project SANDY, grant No.~13N7092) and by the DFG (grant no.
He 2789). They are also grateful to
Henk Taale and the Dutch {\it Ministry of Transport, 
Public Works and Water Management} for supplying the freeway data.

\clearpage
\begin{table}
\caption[]{
\label{tab_par}
Typical values of model parameters used in the non-local, gas-kinetic-based
traffic model to reproduce empirical data of the Dutch freeway A9.}
\begin{center}
\begin{tabular}{|c|c|c|c|c|} \hline
  \,$V_0$\vphantom{$\int_a^b$}  \,&\, $T$   \,&\, $\tau$ 
         \,&\,$\rho_{\rm max}$ \,&\, $\gamma$ \, \\ \hline \hline
  \,110 km/h\vphantom{$\int_a^b$} \,&\, 1.8 s \,&\, 35 s 
  \,&\, 160 veh./km \,&\, 1.2\, \\ \hline
\end{tabular}
\end{center}
\end{table}

\begin{table}[hbt]
\caption{Model parameters of the IDM model used throughout this 
paper.\label{tab:param}}
\begin{center}
\begin{tabular}{|c|c|c|c|c|c|c|c|} \hline
  \,Type\vphantom{$\int_a^b$} \,&\, $v_0$     \,&\, $T$   \,&\, $a$ 
         \,&\,$b$ \,&\, $s_0$     \,&\, $s_1$ \,&\, $l$\, \\ \hline \hline
  \,Cars\vphantom{$\int_a^b$} \,&\, 120 km/h  \,&\, 1.2 s \,&\, 0.8 m/s$^2$ 
     \,&\, 1.25 m/s$^2$ \,&\, 1 m       \,&\, 10 m  \,&\, 5 m\, \\ \hline
  \,Lorries\vphantom{$\int_a^b$} \,&\, 80 km/h \,&\, 1.7 s \,&\, 0.4 m/s$^2$ 
     \,&\, 0.8  m/s$^2$ \,&\, 1 m       \,&\, 10 m  \,&\, 8 m\, \\
\hline
\end{tabular}
\end{center}
\vspace*{-3mm}
\end{table}
\clearpage

\unitlength=0.8mm 

\unitlength=10mm  

\unitlength1.6cm
\clearpage

\clearpage

\clearpage

\begin{thebibliography}{10}

\bibitem{Leu88}
W. Leutzbach, {\it Introduction to the Theory of Traffic Flow},
Springer, Berlin (1988).

\bibitem{book}
D. Helbing, {\em Verkehrsdynamik}, Springer, Berlin (1997). 

\bibitem{traf}  M. Schreckenberg and D. E. Wolf  
(eds.) {\em Traffic and Granular Flow '97}, 
Springer, Singapore (1998).

\bibitem{Bovy} P. H. L. Bovy (ed.) {\em Motorway Traffic Flow Analysis.
New Methodologies and Recent Empirical Findings}, Delft University Press,
Delft (1998).

\bibitem{kerner-sync}
B. S. Kerner and H. Rehborn, 
Experimental properties of phase transitions in traffic flow,
{\em Physical Review Letters} {\bf 49}, 4030--4033 (1997).

\bibitem{helb-nat}
D. Helbing and B. A. Huberman, 
Coherent moving states in highway traffic, 
{\em Nature} {\bf 396}, 738--740 (1998). 

\bibitem{kk-93}
B. S. Kerner and P. Konh{\"a}user, 
Cluster effect in initially homogeneous traffic flow,
{\it Physical Review E} {\bf 48}, R2335--R2338 (1993).

\bibitem{Kue84}
R. D. K{\"u}hne, 
Macroscopic freeway model for dense traffic---{S}top-start 
waves and incident detection. In
{\it Proceedings of the 9th International Symposium on Transportation 
and Traffic Theory} (edited by I. Volmuller and R. Hamerslag), pp.~21--42,
VNU Science Press, Utrecht, The Netherlands (1984).

\bibitem{Kue87}
R. D. K{\"u}hne, 
Freeway speed distribution and acceleration
noise---{C}alculations from a stochastic continuum theory and comparison with
measurements. In {\it Proceedings of the 10th
  International Symposium on Transportation and Traffic Theory}
(edited by N.~H. Gartner and N.~H.~M. Wilson), pp.~119--137,
Elsevier, New York (1987).

\bibitem{kk-94}
B. S. Kerner and P. Konh{\"a}user, 
Structure and parameters of clusters in traffic flow,
{\it Physical Review E} {\bf 50}, 54--83 (1994).

\bibitem{kerner-rehb96-2}
B. S. Kerner and H. Rehborn, 
Experimental properties of complexity in traffic flow,
{\it Physical Review E} {\bf 53}, R4275--R4278 (1996).

\bibitem{HT-sync}
D. Helbing and M. Treiber, 
Gas-kinetic-based traffic model explaining observed hysteretic
phase transition,
{\em Physical Review Letters} {\bf 81}, 3042-3045 (1998).

\bibitem{kerner-rehb96}
B. S. Kerner and H. Rehborn, 
Experimental features and characteristics of traffic jams,
{\it Physical Review E} {\bf 53}, R1297--R1300 (1996).

\bibitem{kerner-chania}
B. S. Kerner, 
Experimental characteristics of traffic flow for evaluation of traffic 
modelling. In {\em Transportation Systems}, Vol. II (edited by M. Papageorgiou
and A. Pouliezos), pp. 793--798, International Federation of Automatic Control,
Chania, Greece (1997).

\bibitem{Ref} R. A. Raub and Pfefer, 
Vehicular flow past incidents involving lane blockage on urban roads:
A preliminary exploration. In 
{\em 77th TRB Annual Meeting}, manuscript no. 004,
Transportation Research Board, Washington, D.C. (1998).

\bibitem{Lee-emp}
H. Y. Lee, H.~W. Lee, and D. Kim, Empirical phase diagram of congested
traffic flow, Preprint {\tt http://xxx.lanl.gov/abs/cond-mat/9905292}.

\bibitem{Treiterer}
J. Treiterer and J. A. Myers, The hysteresis phenomenon in traffic
flow.
In {\em Proc. 6th Int. Symp. on Transportation and Traffic Theory}
(edited by D.~J. Buckley), pp. 13--38, Elsevier, New York (1974). 

\bibitem{OCT}
C. F. Daganzo, M. J. Cassidy, and R. L. Bertini, Some traffic features at
freeway bottlenecks, {\it Transportation Research B} {\bf 33}, 25--42 (1999). 

\bibitem{Kerner-wide}
B. S. Kerner, Experimental features of self-organization in traffic flow,
{\it Physical Review Letters} {\bf 81}, 3797--3800 (1998). 

\bibitem{GKT}
M. Treiber, A. Hennecke, and D. Helbing, Derivation, properties, and
simulation of a gas-kinetic-based, non-local traffic model, 
{\it Physical Review E} {\bf 59}, 239--253 (1999). 

\bibitem{coexist}
M. Treiber and D. Helbing, Explanation of observed features of
self-organization in traffic flow. Preprint 
{\tt http://xxx.lanl.gov/abs/cond-mat/9901239}.

\bibitem{Phase}
D. Helbing, A. Hennecke, and M. Treiber, Phase diagram of traffic states in
the presence of inhomogeneities. {\it Physical Review Letters} {\bf 82},
4360--4363 (1999).

\bibitem{LW} M. J. Lighthill and G. B. Whitham, 
On kinematic waves: {II. A} theory of traffic on long crowded roads,
{\it Proceedings of the Royal Society A} {\bf 229}, 317--345 (1955).

\bibitem{Richards} P. I. Richards, 
Shock waves on the highway,
{\it Operations Research} {\bf 4}, 42--51 (1956).

\bibitem{New} G. F. Newell,
A simplified theory of kinematic waves in highway traffic,
{\it Transportation Research B} {\bf 27}, 281--313 (1993).

\bibitem{Dag} C. F. Daganzo, 
The cell transmission model: 
{A} dynamic representation of highway traffic consistent 
with the hydrodynamic theory,
{\it Transportation Research B} {\bf 28}, 269--287 (1994).

\bibitem{Dag2}
D. F. Daganzo, 
The cell transmission model, {P}art {II}: {N}etwork traffic,
{\it Transportation Research B} {\bf 29}, 79--93 (1995).

\bibitem{Hil95}
M. Hilliges and W. Weidlich, 
A phenomenological model for dynamic traffic flow in networks,
{\it Transportation Research B} {\bf 29}, 407--431 (1995).

\bibitem{Lebacque} J. P. Lebacque, 
A finite acceleration scheme for first order macroscopic traffic flow models.
In {\em Transportation Systems}, Vol. II (edited by M. Papageorigiou
and A. Pouliezos), pp. 815--820, International Federation of Automatic Control,
Chania, Greece (1997).

\bibitem{Payne}
H. J. Payne, Models of freeway traffic and control. In
{\it Mathematical Models of Public Systems}, Vol. 1 (edited by G.~A. Bekey),
pp.~51--61, Simulation Council, La Jolla, CA (1971).

\bibitem{Papa} M. Papageorgiou, {\em Applications of Automatic Control 
Concepts to Traffic Flow Modeling and Control}, Springer, Berlin (1983).

\bibitem{Cremer} M. Cremer, 
{\em Der Verkehrsflu{\ss} auf Schnellstra{\ss}en},
Springer, Berlin, 1979. 

\bibitem{phillips-kin}
W. F. Phillips, 
A kinetic model for traffic flow with continuum implications,
{\it Transportation Planning and Technology} {\bf 5}, 131--138 (1979).

\bibitem{Lee}
H. Y. Lee, H.~W. Lee, D., and Kim, Origin of synchronized traffic flow on
highways and its dynamic phase transition, {\it Phys. Rev. Lett.} 
{\bf 81}, 1130--1133 (1998).

\bibitem{Whitham}
G. B. Whitham, {\it Linear and Nonlinear Waves},
Wiley, New York (1974).

\bibitem{Dag95}
C. F. Daganzo,
Requiem for second-order fluid approximations of traffic flow,
{\it Transportation Research B} {\bf 29}, 277--286 (1995).

\bibitem{num} D. Helbing and M. Treiber,  
Numerical simulation of macroscopic traffic equations,
{\em Computing in Science and Engineering} {\bf 1}, 89--99 (1999).

\bibitem{bando}
M. Bando, K. Hasebe, K. Nakanishi, A. Nakayama, A. Shibata, and Y.
Sugiyama, 
Phenomenological study of dynamical model of traffic flow,
{\em Journal de Physique I France} {\bf 5}, 1389--1399 (1995). 

\bibitem{zellauto}
D. Helbing and M. Schreckenberg, 
Cellular automata simulating experimental properties of traffic flow,
{\em Physical Review E} {\bf 59}, R2505--R2508 (1999).

\bibitem{pre}
D. Helbing, 
Gas-kinetic derivation of Navier-Stokes-like traffic equations,
{\em Physical Review E} {\bf 53}, 2366--2381 (1996).

\bibitem{physa}
D. Helbing, 
Derivation and empirical validation of a refined traffic flow model,
{\em Physica A} {\bf 233}, 253--282 (1996). 

\bibitem{prigogine}
I. Prigogine and R. Herman, 
{\em Kinetic Theory of Vehicular Traffic}, Elsevier, New York (1971). 

\bibitem{paveri-fontana}
S.-L. Paveri-Fontana, 
On {B}oltzmann-like treatments for traffic flow.
{A} critical review of the basic model and an alternative proposal for dilute
traffic analysis,
{\it Transportation Research} {\bf 9}, 225--235 (1975).

\bibitem{nelson}
P. Nelson, 
A kinetic model of vehicular traffic and its associated
bimodal equilibrium solutions,
{\it Transport Theory and Statistical Physics} {\bf 24}, 383--409 (1995).

\bibitem{wagner-96}
C. Wagner {\it et~al.}, 
Second order continuum traffic flow model,
{\it Physical Review E} {\bf 54},  5073--5085 (1996).  

\bibitem{klar-97}
A. Klar and R. Wegener, 
Enskog-like kinetic models for vehicular traffic,
{\em Journal of Statistical Physics} {\bf 87}, 91--114 (1997).

\bibitem{alex}
P. Nelson and A. Sopasakis,
The Prigogine-Herman kinetic model predicts widely scattered traffic 
flow data at high concentrations, 
{\em Transportation Research B} {\bf 32}, 589--604 (1998).

\bibitem{hoog}
S. Hoogendoorn and P.~H.~L. Bovy,  
A macroscopic multi-lane multi-class traffic flow model,
Paper submitted to
the 79th TRB Annual Meeting 1999 in Washington (1998).

\bibitem{kw-98a}
A. Klar and R. Wegener,
A hierarchy of models for multilane traffic I: Modeling,
{\em SIAM Journal on Applied Mathematics} {\bf 59}(3), 983--1001 (1999).

\bibitem{kw-98b}
A. Klar and R. Wegener, 
A hierarchy of models for multilane traffic II: Numerical investigations,
{\em SIAM Journal on Applied Mathematics} {\bf 59}(3), 1002-1011 (1999).

\bibitem{parisi}
D. Helbing, 
From microscopic to macroscopic traffic models. In
{\em A Perspective Look at Nonlinear Media. From Physics to
  Biology and Social Sciences} (edited by J. Parisi, S. C. M\"uller, and
W. Zimmermann), pp. 122--139, Springer, Berlin, 1998. 

\bibitem{drygran}
D. Helbing and M. Treiber, 
Enskog equations for traffic flow evaluated up to Navier-Stokes order,
{\em Granular Matter} {\bf 1}, 21--31 (1998).

\bibitem{mul-HS} 
V. Shvetsov and D. Helbing, 
Macroscopic dynamics of multi-lane traffic,
{\em Physical Review E} {\bf 59}, 6328--6339 (1999). 

\bibitem{emp-Hel}
D. Helbing,
Empirical traffic data and their implications for traffic modeling,
{\em Physical Review E} {\bf 55}, R25-R28 (1997).

\bibitem{GKT-scatter}
M. Treiber and D. Helbing, Macroscopic simulation of widely scattered
synchronized traffic states, {\it Journal of Physics A: 
Mathematical and General} {\bf 32} 
L17--L23 (1999).

\bibitem{fund}
D. Helbing,
Fundamentals of traffic flow,
{\em Physical Review E} {\bf 55}, 3735--3738 (1997).

\bibitem{kerner-ramp}
B. S. Kerner, P. Konh{\"a}user, and M. Schilke, 
Deterministic spontaneous
appearance of traffic jams in slightly inhomogeneous traffic flow,
{\it Physical Review E} {\bf 51}, R6243--R6246 (1995).

\bibitem{kerner-dipole}
B. S. Kerner, P. Konh{\"a}user, and M. Schilke, 
``Dipole-layer'' effect in dense traffic flow,
{\em Physics Letters A} {\bf 215}, 45--56 (1996). 

\bibitem{Cross}
M. C. Cross and P. C. Hohenberg, 
Pattern formation outside of equilibrium,
{\it Review of Modern Physics} {\bf 65}, 851--1112 (1993). 

\bibitem{ames}
W. F. Ames, {\it Nonlinear Partial Differential Equations in
Engineering}, Vols. I \& II, Academic Press, New York (1965).

\bibitem{numrep}
W. H. Press, S. A. Teukolsky, W. T.  Vetterling, and B. P. Flannery, 
{\it Numerical Recipes in C: The Art of Scientific Computing},
Cambridge University Press, Cambridge (1992).

\bibitem{Daganzo} C. F. Daganzo, M. J. Cassidy, and R. L. Bertini, 
Possible explanations of phase transitions in highway traffic,
{\em Transportation Research A} {\bf 33}, 365--379 (1999).

\bibitem{infl} C. F. Daganzo, 
The nature of freeway gridlock and how to prevent it. In
{\em Proc. 13th Int. Symp. on Transportation and Traffic Theory}
(edited by J. B. Lesort), pp. 629--646, 
Pergamon-Elsevier, New York (1996).

\bibitem{ML} 
D. Helbing, 
Modeling multi-lane traffic flow with queuing effects,
{\em Physica A} {\bf 242}, 175--194 (1997).

\bibitem{HT-Science}
D. Helbing and M. Treiber, 
Jams, waves, and clusters,
{\em Science} {\bf 282}, 2001--2003 (1998).

\bibitem{Lee99}
H.~Y. Lee, H.~W. Lee, and D. Kim, 
Dynamic states of a continuum traffic equation with on-ramp,
{\em Physical Review E} {\bf 59}, 5101--5111  (1999).

\bibitem{opus}
M. Treiber, A. Hennecke, and D. Helbing,
Congested traffic states in empirical observations and microscopic
simulations, preprint {\tt http://arXiv.org/abs/cond-mat/0002177},
submitted to {\em Physical Review E} (2000).

\bibitem{transresb}
D. Helbing, A. Hennecke, V. Shvetsov, and M. Treiber 
Macroscopic traffic simulation based on a gas-kinetic, non-local traffic
model, {\em Transportation Research B}, in print (2000).

\bibitem{Reuschel}
A. Reuschel, Fahrzeugbewegungen in der Kolonne, {\it {\"O}sterreichisches
Ingenieur-Archiv} {\bf 4}, 193--215 (1950).

\bibitem{Newell}
G. F. Newell, Nonlinear effects in the dynamics of car following,
{\it Operations Research} {\bf 9}, 209--229 (1961).

\bibitem{Bando}
M. Bando, K. Hasebe, A. Nakayama, A. Shibata, and Y. Sugiyama,
Dynamical model of traffic congestion and numerical simulation, 
{\it Physical Review E} {\bf 51}, 1035--1042 (1995). 

\bibitem{Gipps81}
P. G. Gipps, A behavioural car-following model for computer simulation,
{\it Transportation Research B} {\bf 15}, 105--111 (1981). 

\bibitem{Krauss-Diss}
S. Krau{\ss}, {\it Microscopic Modelling of Traffic Flow}, 
PhD thesis, DLR, FB 98-08, Cologne, (1998).

\bibitem{Tilch-GFM}
D. Helbing and B. Tilch, Generalized force model of traffic
dynamics, {\it Phys. Rev. E} {\bf 58}, 133--138 (1998).

\bibitem{Wiedemann}
R. Wiedemann, {\it Simulation des Stra{\ss}enverkehrsflusses},
Institut f\"ur Ver\-kehrs\-we\-sen, Universit\"at Karlsruhe (1974).

\bibitem{MITSIM}
See internet page {\tt
http://hippo.mit.edu/products/mitsim/main.html}.

\bibitem{Applet-engl}
Interactive simulations of the multi-lane IDM are available at {\tt
http://www.uni-stuttgart.de/treiber/MicroApplet/}.

\bibitem{Sollacher-control}
R. Sollacher and H. Lenz, Nonlinear control of stop-and-go traffic.
In {\em Traffic and Granular Flow '99} (edited by D. Helbing,
H. J. Herrmann, M. Schreckenberg, and D. E. Wolf), Springer,
Berlin, 2000.

\end{thebibliography}
\end{document}